\documentclass[structabstract]{aa}  
\usepackage{graphicx}
\usepackage{txfonts}
\begin{document}
   \title{Dwarf Spheroidals in the M81 Group -- Metallicity Distribution
    Functions and Population Gradients}

   \author{S.~Lianou 
              \inst{1} 
              \fnmsep\thanks{Fellow of the Heidelberg Graduate School
                             of Fundamental Physics (HGSFP) and member
                             of the International Max Planck Research
                             School (IMPRS) for Astronomy \& Cosmic
                             Physics at the University of Heidelberg}
          \and
           E.~K.~Grebel 
              \inst{1} 
          \and 
           A.~Koch 
              \inst{2}
          }

           \institute{Astronomisches Rechen-Institut, Zentrum f\"ur
                  Astronomie der Universit\"at Heidelberg,
                  M\"onchhofstrasse 12-14, D-69120 Heidelberg,
                  Germany\\  
                  \email{lianou@ari.uni.heidelberg.de}; \email{grebel@ari.uni.heidelberg.de}  
            \and
                  University of Leicester, Department of Physics and
                  Astronomy, University Road, LE1 7RH Leicester, UK\\
                  \email{ak326@astro.le.ac.uk}
                      }

   \date{Received September 28, 2009; accepted February 25, 2010}

% \abstract{}{}{}{}{} 
% 5 {} token are mandatory
 
  \abstract
   {}
   {We study the dwarf spheroidal galaxies in the nearby M81 group in
     order to construct their photometric metallicity distributions
     and to investigate the potential presence of population
     gradients. We select all the dwarf spheroidals with available
     Hubble Space Telescope\,/\,Advanced Camera for Surveys archival
     observations, nine in total.}
   {We interpolate isochrones so as to assign a photometric
     metallicity to each star within a selection box in the
     color-magnitude diagram of each dwarf galaxy. We assume that the
     dwarf spheroidals contain mainly an old stellar population. In
     order to search for metallicity gradients, we examine the spatial
     distribution of two stellar populations that we separate
     according to their metallicities.}
   {As a result, we present the photometric metallicity distribution
     functions, the cumulative histograms and smoothed density maps of
     the metal-poor and metal-rich stars as well as of the
     intermediate-age stars.}
   {From our photometric data we find that all the dwarf spheroidals
     show a wide range in metallicities, with mean values that are
     typical for old and metal-poor systems, with the exception of one
     dwarf spheroidal, namely IKN. Some of our dwarf spheroidals
     exhibit characteristics of transition-type dwarfs. Compared to
     the Local Group transition type dwarfs, the M81 group ones appear
     to have mean metallicity values slightly more metal-rich
     at a given luminosity. All the dwarf spheroidals
     considered here appear to exhibit either population gradients or
     spatial variations in the centroids of their metal-poor and
     metal-rich population. In addition, there are luminous AGB stars
     detected in all of them with spatial distributions suggesting
     that they are well mixed with the old stars.}


  \titlerunning{M81 group dSphs -- MDFs and population gradients}
  \maketitle

\section{Introduction}
\label{sec:introduction}

    The dwarf galaxies within our Local Group (LG) have been the
    subject of intensive spectroscopic and photometric observations in
    different wavelength regimes and thus are well studied
    objects. Their study has been facilitated by the proximity of
    these dwarfs so that individual stars may even be resolved down to
    the main sequence turn-off, depending on their distance. Thus,
    extending the studies to dwarf galaxies in nearby groups with
    different environment and comparing their properties are of great
    importance in order to understand the main drivers of their
    evolution. In addition, the derived properties can provide a way
    to constrain models of galaxy formation as well as chemical
    evolutionary models

    In this respect, the \object{M\,81 group} is an interesting target: despite
    several differences, it bears close resemblance to our LG. The
    similarity of the M\,81 group to our LG lies in its binary
    structure (Karachentsev et al.~\cite{sl_m81distances}), while its
    difference is mainly due to the recent interactions between its
    dominant consituents as revealed by the formation of tidal tails
    and bridges detected in HI observations (Appleton, Davies \&
    Stephenson \cite{sl_appleton}; Yun, Ho \& Lo \cite{sl_yun}). With
    a mean distance of $\sim$3.7~Mpc (Karachentsev et
    al.~\cite{sl_m81distances}), the M\,81 group is one of the nearest
    groups to our own LG. It consists of about 40 dwarfs of both
    early-type and late-type, with the addition of 12 recently
    discovered dwarf candidates (Chiboucas, Karachentsev \& Tully
    \cite{sl_chiboucas}).

    Here we focus on the dwarf spheroidal galaxies (dSphs) in the
    M\,81 group with available Hubble Space Telescope
    (HST)\,/\,Advanced Camera for Surveys (ACS) archival data. The
    dSphs are objects with low surface brightness and poor in gas
    content. For a summary of their properties we refer to Grebel,
    Gallagher \& Harbeck (\cite{sl_grebel}; and references
    therein). We use their color-magnitude diagrams (CMDs) to derive
    the photometric metallicity distribution functions (MDFs) and
    search for the potential presence of population gradients in the
    M\,81 group dSphs.

    The use of the CMD to infer the star formation histories and MDFs
    is a very powerful tool. For nearby groups at distances, where
    individual red giants are beyond the reach of spectroscopy even
    with 8-10~m class telescopes, CMDs are the best means to derive
    evolutionary histories. With the use of HST observations of
    adequate depth, the upper part of the red giant branch (RGB) can
    be resolved into single stars. Many studies have derived the
    photometric MDFs of distant LG dSphs (for example Cetus by
    Sarajedini et al.~\cite{sl_sarajedini}; And\,VI and And\,VII by
    Grebel \& Guhathakurta \cite{sl_grebel99}) from their CMDs. A
    similar work to derive the photometric MDFs for dwarf galaxies in
    nearby groups has not been done so far.  

    The search for radial population gradients in LG dwarf galaxies
    has been favoured by the fact that the resolved stellar
    populations reach the horizontal branch or extend even below the
    main-sequence turn-off depending on the distance of the dwarf,
    permitting one to use a variety of different stellar
    tracers. There are several studies for population gradients in the
    LG dwarfs and as an example of such studies we refer to the work
    done by Hurley-Keller, Mateo \& Grebel
    (\cite{sl_hurley-keller99}), Harbeck et al.~(\cite{sl_harbeck}),
    Battaglia et al.~(\cite{sl_battaglia06}) (photometric) and Tolstoy
    et al.~(\cite{sl_tolstoy}), Koch et al.~(\cite{sl_koch06})
    (spectroscopic). There is not any study so far searching for
    population gradients in nearby group dwarf galaxies.  
 
    This paper is structured as follows. In \S2 we present the
    observations, in \S3 we show our results, in \S4 we discuss our
    main findings and in \S5 we present our conclusions.

 \section{Observations}
 \label{sec:observations}

    We use HST\,/\,ACS archival data that were retrieved
%
%%%%% TABLE 1 %%%%%%%%%%%%%%%%%%%%% Two column Table %%%%%%%%%%%%%%%%%%%%%%%%%%%%%%%
\begin{table*}
     \begin{minipage}[t]{\textwidth}
      \caption{Log of Observations.}
      \label{table1} 
      \centering
      \renewcommand{\footnoterule}{}  % to avoid a line before footnotes
      \begin{tabular}{l c c c c c }   % 6 columns  
\hline\hline
    Galaxy      	&RA~(J2\,000.0)   &Dec~(J2\,000.0)    &Program~ID~/~PI            &ACS\,/\,WFC Filters  &Exposure time           \\ 
                        &                 &                   &                           &                     &(s)                     \\
    (1)                 &(2)              &(3)                &(4)                        &(5)                  &(6)                     \\ 
\hline 

    KDG 61              &09~57~03.10      &$+$68~35~31.0      &GO~9\,884~/~Armandroff     &F606W\,/\,F814W      &8\,600~/~9\,000         \\

    KDG 64	   	&10~07~01.90	  &$+$67~49~39.0      &...                        &...                  &...                     \\

    DDO 71 	   	&10~05~06.40	  &$+$66~33~32.0      &...                        &...                  &...                     \\

    F12D1	   	&09~50~10.50	  &$+$67~30~24.0      &...                        &...                  &...                     \\

    F6D1	   	&09~45~10.00	  &$+$68~45~54.0      &...                        &...                  &...                     \\

    \hline 
    
    HS\,117	        &10~21~25.20	  &$+$71~06~51.0      &SNAP~9\,771~/~Karachentsev &F606W\,/\,F814W      &1\,200~/~900            \\

    IKN	   	        &10~08~05.90	  &$+$68~23~57.0      &...                        &...                  &...                     \\

\hline 
   
    DDO\,78	        &10~26~28.00	  &$+$67~39~35.0      &GO~10\,915~/~Dalcanton     &F475W\,/\,F814W      &2\,274~/~2\,292         \\

    DDO\,44  	        &07~34~11.50	  &$+$66~52~47.0      &...                        &...                  &2\,361~/~2\,430         \\

\hline

\end{tabular} 
\footnotetext{Note.-- Units of right ascension are hours,
                      minutes, and seconds, and units of declination
                      are degrees, arcminutes and arcseconds.}%
\end{minipage}
\end{table*}
%%%%%%%%%%%%%%%%%%%%%%%%%%%%%%%%%%%%%%%%%%%%%%%%%%%%%%%%%%%%%%%%%%%%%%%%%%%%%%%%%%%%
%
    through the Multimission Archive at STScI (MAST). The details of
    the datasets used are listed in Table~\ref{table1}, where the
    columns show: (1) the galaxy name, (2) and (3) equatorial
    coordinates of the field centers (J2000.0), (4) the number of the
    Program ID and the PI, (5) the ACS\,/\,WFC filters used, and (6)
    the total exposure time for each filter.

    The data reduction was carried out using Dolphot, a modified
    version of the HSTphot photometry package (Dolphin
    \cite{sl_dolphin2000}) developed specifically for ACS point source
    photometry. The reduction steps followed are the ones described in
    the ACS module of the Dolphot manual. In the Dolphot output
    photometric catalogue, only objects with $S/N >$~5 and ``type''
    equal to 1, which means ``good stars'', were allowed to enter the
    final photometric catalogue. The ``type'' is a Dolphot parameter
    that is used to distinguish objects that are classified as ``good
    stars'', ``elongated objects'', ``too sharp objects'' and so
    on. After this first selection, quality cuts were applied so as to
    further clean the photometric catalogue. These cuts were based on
    the distributions of the sharpness and crowding parameters, as
    suggested in the Dolphot manual and also in Williams et
    al.~(\cite{sl_angst}). Guided by these distributions, we use for
    the sharpness parameter the restriction of
    $|sharpness_{filter}+sharpness_{F814W}|<\ \alpha$, where
    (1.0~$<\alpha<$~1.5) depending on the dSph, and for the crowding
    parameter the requirement
    ($Crowding_{filter}~+~Crowding_{F814W})<$~1.0, where ``filter''
    corresponds to either the F606W or the F475W filter. These
    selections were made so as to ensure that only stellar objects
    have entered our final photometric catalogue. The number of stars
    recovered after applying all the photometric selection criteria
    are listed in Table~\ref{table2}, column (3). 

    The photometry obtained with Dolphot provides magnitudes in both
    the ACS\,/\,WFC and the Landolt UBVRI photometric systems using
    the transformations provided by Sirianni et
    al.~(\cite{sl_sirianni}) for the UBVRI system. In the analysis
    presented throughout this work, we chose to use the ACS\,/\,WFC
    filter system. Therefore, if we use data from the literature
    computed in the UBVRI photometric system, we transform them to the
    ACS\,/\,WFC system. There are two cases where this is
    necessary. The first case is the extinction. The galactic
    foreground extinction in the V-band and I-band, $A_{I}$ and
    $A_{V}$, taken from Schlegel, Finkbeiner \& Davis
    (\cite{sl_schlegel}) through NED, are transformed into the
    ACS\,/\,WFC system. For the transformation, we use the
    corresponding extinction ratios $A(P)\,/\,E(B-V)$ for a G2 star,
    where $A(P)$ corresponds to the extinctions in the filters F814W
    and F606W (or F475W), which are provided by Sirianni et
    al.~(\cite{sl_sirianni}; their Table 14). We note that the
    assumption of a largely color-independent reddening for the RGB is
    justified since theoretical models indicate that the expected
    effect of the change of color accross the RGB amounts to at most
    0.01 in $E(V-I)$ for our data (see Grebel \& Roberts
    \cite{sl_grebel95}). We multiply these extinction ratios with the
    $E(B-V)$, in order to finally get the extinctions in the ACS
    filters. The transformed values, $A_{F814W}$ and $A_{F606W}$ (or
    $A_{F475W}$), are listed in Table~\ref{table2}, columns (6) and
    (7) respectively.

    The second case is the I-band tip of the RGB (TRGB), which we
    transform to the F814W-band TRGB in the following way. As already
    mentioned, Dolphot provides the magnitudes both in the
    instrumental ACS\,/\,WFC system and in the transformed
    UBVRI. Thus, in the range of magnitudes near the I-band TRGB, we
    compute the difference in magnitudes between the I-band and
    F814W-band. This difference is 0.01~mag for all the dSphs except
    for DDO\,44 and DDO\,78, where the difference is $-$0.015~mag. The
    F814W-band TRGB is then equal to the sum of this difference and
    the I-band TRGB. We confirm further more this approach of
    estimating the F814W-band TRGB by applying a Sobel-filtering
    technique to the luminosity function of some of the dSphs (Lee,
    Freedman \& Madore \cite{sl_lee}; Sakai, Madore \& Freedman
    \cite{sl_sakai}) estimating the location of the F814W-band
    TRGB. We find that these approaches give values that are in good
    agreement, with a mean difference between them of the order of
    0.05~mag.

%%%%% TABLE 2 %%%%%%%%%%%%%%%%%%%%% Two column Table %%%%%%%%%%%%%%%%%%%%%%%%%%%%%%%
\begin{table*}
\begin{minipage}[t]{\textwidth}
\caption[]{Global Properties (see text for references).}
\label{table2} 
\centering
\renewcommand{\footnoterule}{}  % to avoid a line before footnotes
\begin{tabular}{l c c c c c c c c c}     % 10 columns  

\hline\hline
  Galaxy    &Type         &$N_{*}$    &$M_V$      &$I_{TRGB}$        &$A_{F814W}$  &$A_{F606W}$\footnote{or $A_{F475W}$ in the case of DDO\,44 and DDO\,78}
                                                                                            &$(m-M)_{O}$       &$R$     &$r_{eff}$         \\          
            &             &           &(mag)      &(mag)            &(mag)       &(mag)     &(mag)             &(kpc)   &$(\prime\prime)$ \\           
   (1)      &(2)          &(3)        &(4)        &(5)              &(6)         &(7)       &(8)               &(9)     &(10)             \\           
                                                                                                                               
\hline                                                                                                                         
                                                                                                                               
  KDG\,61  &dIrr~/~dSph   &$53\,543$  &$-13.87$    &$23.86\pm0.15$   &$0.131$     &$0.202$   &$27.78\pm0.15$    &$44$   &$48$             \\           
                                                                                                                               
  KDG\,64  &dIrr~/~dSph   &$38\,012$  &$-13.43$    &$23.90\pm0.15$   &$0.099$     &$0.152$   &$27.84\pm0.15$    &$126$  &$28$             \\           
                                                                                                                               
  DDO\,71  &dIrr~/~dSph   &$37\,291$  &$-13.22$    &$23.83\pm0.15$   &$0.173$     &$0.267$   &$27.72\pm0.15$    &$211$  &$59$             \\           
                                                                                                                               
  F12D1    &dSph          &$39\,519$  &$-12.84$    &$23.95\pm0.15$   &$0.263$     &$0.404$   &$27.71\pm0.15$    &$181$  &$31$             \\           
                                                                                                                               
  DDO\,78  &dSph          &$21\,073$  &$-12.83$    &$23.85\pm0.15$   &$0.040$     &$0.079$   &$27.85\pm0.15$    &$223$  &$38$             \\           
                                                                                                                               
  DDO\,44  &dSph          &$19\,357$  &$-12.56$    &$23.55\pm0.15$   &$0.075$     &$0.149$   &$27.52\pm0.15$    &$901$  &$28$             \\           
                                                                                                                               
  IKN	   &dSph          &$14\,600$  &$-11.51$    &$23.94\pm0.15$   &$0.111$     &$0.171$   &$27.87\pm0.18$    &$110$  &...              \\           
                                                                                                                               
  F6D1	   &dSph          &$14\,260$  &$-11.46$    &$23.77\pm0.14$   &$0.144$     &$0.222$   &$27.66\pm0.17$    &$218$  &$32$             \\
                                                                                                                               
  HS\,117  &dIrr~/~dSph   &$4\,596$   &$-11.31$    &$24.16\pm0.15$   &$0.210$     &$0.323$   &$27.99\pm0.18$    &$204$  &$29$             \\           

\hline
 
\end{tabular}
\end{minipage}
\end{table*}
%%%%%%%%%%%%%%%%%%%%%%%%%%%%%%%%%%%%%%%%%%%%%%%%%%%%%%%%%%%%%%%%%%%%%%%%%%%%%%%%%%%%
%
    We list the global properties of the present dSph sample in
    Table~\ref{table2}. The columns contain the following information:
    (1) the galaxy name, (2) the galaxy type, (3) the number of stars
    detected after applying all the photometric selection criteria,
    (4) the visual absolute magnitude $M_{V}$ of each galaxy adopted
    from Karachentsev et al.~(\cite{sl_wfpc2data},
    \cite{sl_f6d1trgb}), Alonso-Garcia, Mateo \& Aparicio
    (\cite{sl_garcia}), Georgiev et al.~(\cite{sl_georgiev}), (5) the
    I-band TRGB adopted from Karachentsev et al.~(\cite{sl_acsdata},
    \cite{sl_f6d1trgb}, \cite{sl_wfpc2data}, \cite{sl_ddo44trgb}), (6)
    and (7) the foreground extinction derived by us for the
    ACS\,/\,WFC filters F814W, F606W and F475W, as described in
    Sec.~\ref{sec:observations}, (8) the true distance moduli adopted
    from Karachentsev et al.~(\cite{sl_acsdata}, \cite{sl_f6d1trgb},
    \cite{sl_wfpc2data}, \cite{sl_ddo44trgb}), (9) the deprojected
    distance of the dSphs from the M\,81 galaxy, R, adopted from
    Karachentsev et al.~(\cite{sl_m81distances}), (10) the effective
    radius, $r_{eff}$ adopted from Sharina et al.~(\cite{sl_sharina})
    and Karachentseva et al.~(\cite{sl_karreff}). The dSphs in
    Table~\ref{table2} are sorted according to their $M_{V}$ value.  

    Finally, the pixel scale of the ACS\,/\,WFC is 0.05\arcsec with a
    field of view of 202\arcsec\,$\times$\,202\arcsec or
    4\,096\,$\times$\,4\,096~pixels. Thus for the mean distance of
    $\sim$3.7~Mpc of the M\,81 group (Karachentsev et
    al.~\cite{sl_m81distances}) this field of view corresponds to
    3.6~kpc\,$\times$\,3.6~kpc, or simply 1~pixel corresponds to
    roughly 1~pc.

\section{Results}
\label{sec:results}

    \subsection{Color-Magnitude Diagrams}
    \label{sec:cmds}

%%%%% FIGURE 1 - CMDs %%%%%%%%%%%%% Two column figure (place early!) %%%%%%%%%%%%%%%
 \begin{figure*}
    \centering
       \includegraphics[width=16cm,clip]{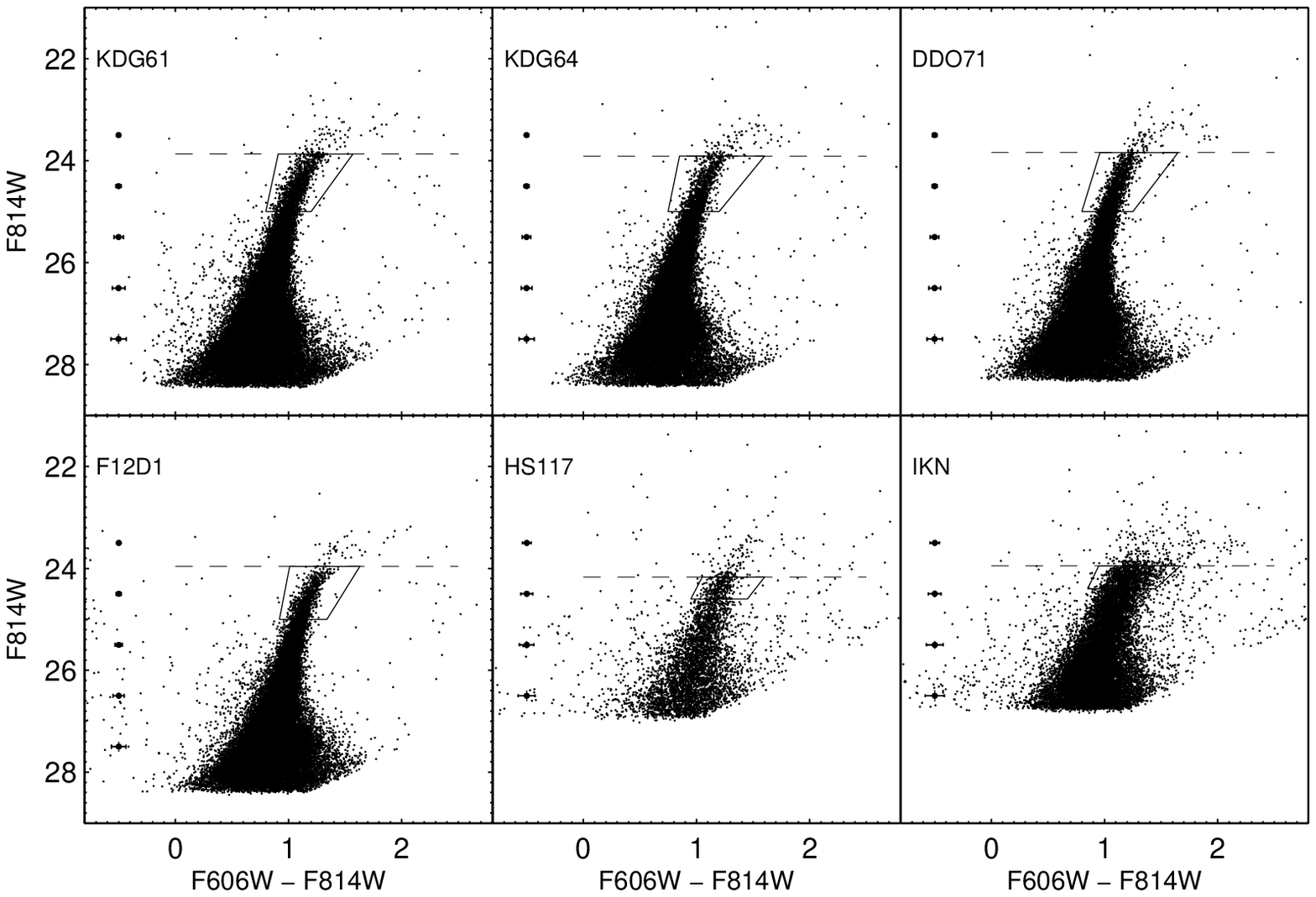}
       \includegraphics[width=16cm,clip]{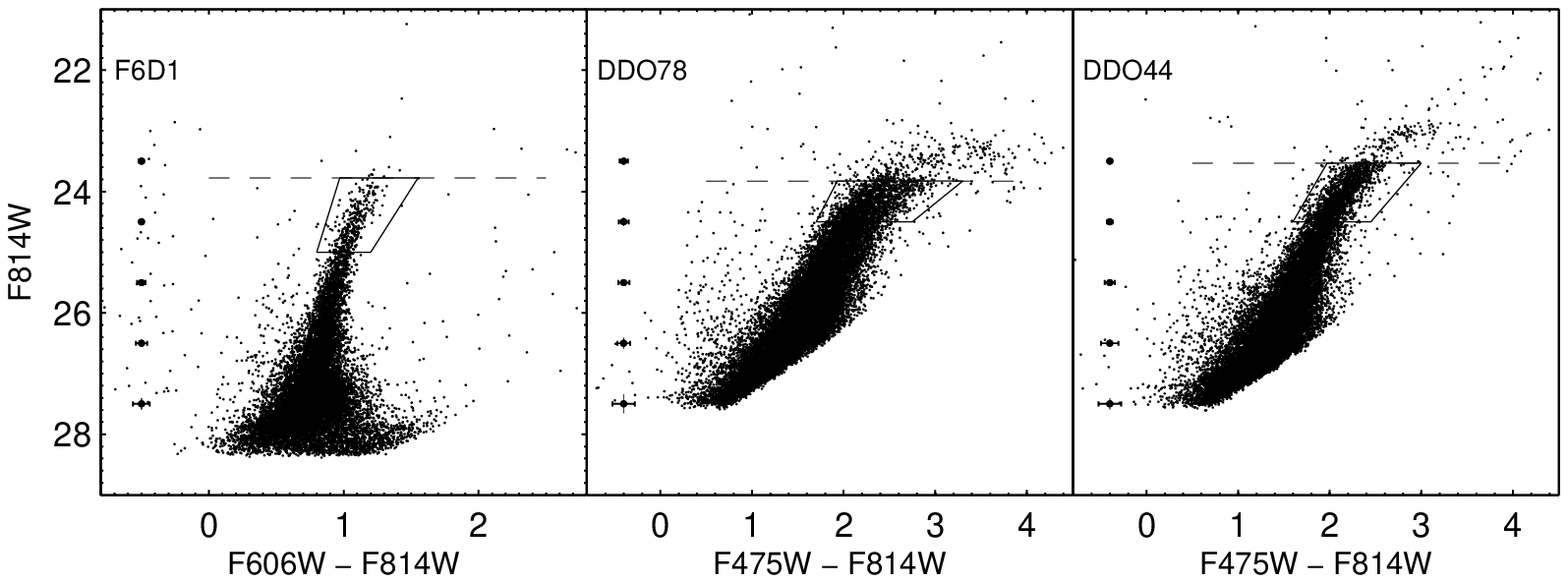}
       \caption{Color-magnitude diagrams for the nine dSphs. The
                horizontal dashed lines show the location of the
                TRGB. The crosses on the left hand side correspond to
                the photometric errors as derived from artificial star
                tests. The boxes enclose the stars for which we derive
                the photometric metallicities.}  
       \label{sl_figure1}%
\end{figure*}
%%%%%%%%%%%%%%%%%%%%%%%%%%%%%%%%%%%%%%%%%%%%%%%%%%%%%%%%%%%%%%%%%%%%%%%%%%%%%%%%%%%%
%
    We show the CMDs of the nine dSphs in Fig.~\ref{sl_figure1}, where
    we note the difference in the x-axis. The proximity of the M\,81
    group and the depth of the observations allow us to resolve the
    upper part of the RGB into individual stars. The most prominent
    feature seen in our CMDs is the RGB. We note the presence of stars
    above the TRGB, which is indicated in Fig.~\ref{sl_figure1} with a
    dashed line. These stars are most likely luminous AGB stars, which
    indicate the presence of stellar populations in an age range from
    1~Gyr up to less than 10~Gyr. In addition, some of the dSphs
    appear to have bluer stars that probably belong to a younger main
    sequence. The dwarfs in our sample are classified as dSphs
    (Karachentsev et al.~\cite{sl_cng}), with the exception of
    KDG\,61, KDG\,64, DDO\,71, and HS\,117 which are classified as
    transition-types (dIrr\,/\,dSph) as they have detectable HI
    content (Huchtmeier \& Skillman \cite{sl_huch}; Boyce et
    al.~\cite{sl_hiblind}) or $H\alpha$ emission (Karachentsev \&
    Kaisin \cite{sl_m81halpha}; Karachentsev et
    al.~\cite{sl_acsdata}).

%%%%% FIGURE 2 - MDFs %%%%%%%%%%%%% Two column figure (place early!) %%%%%%%%%%%%%%%
 \begin{figure*}
  \centering
       \includegraphics[width=17cm,clip]{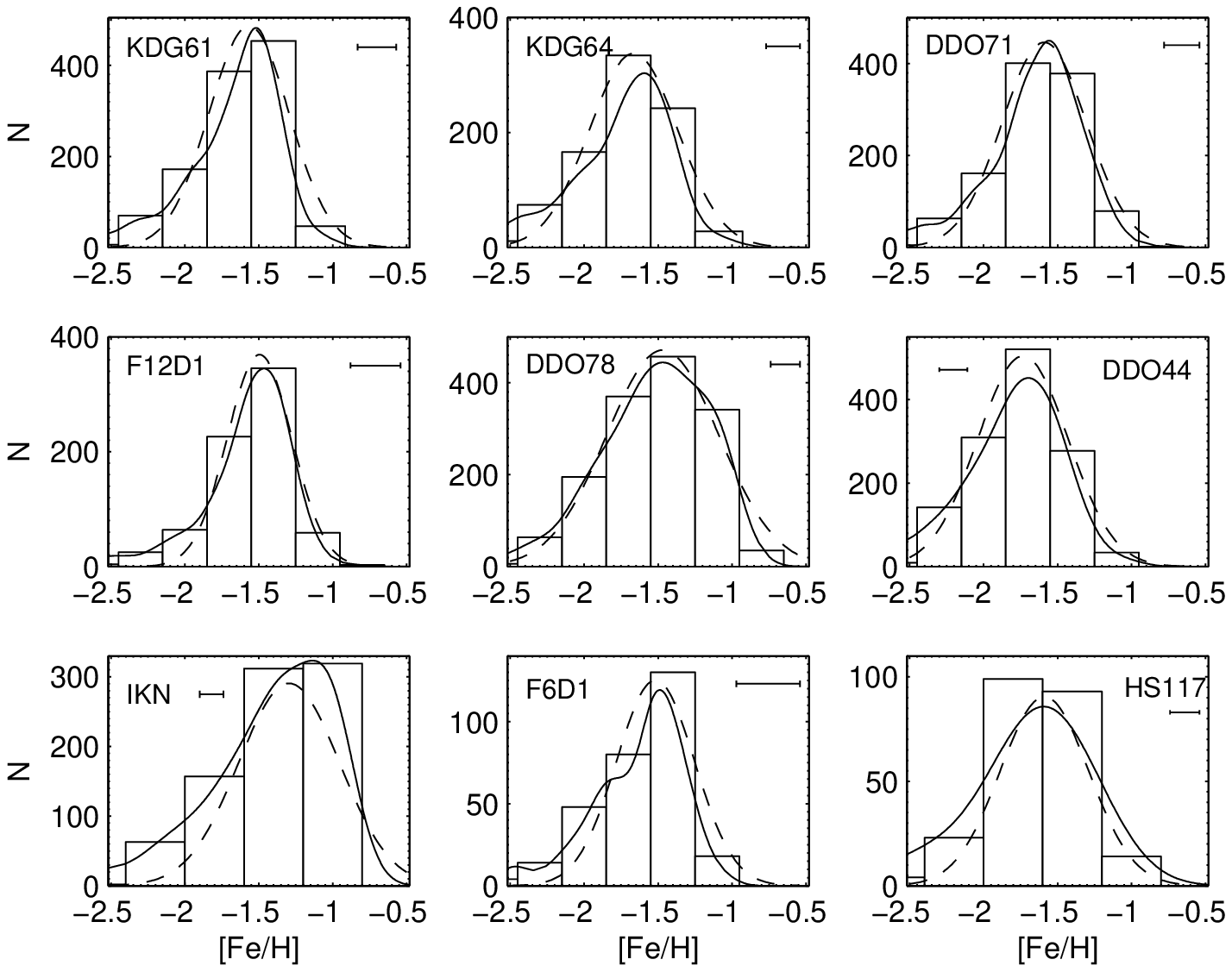}
       \caption{Metallicity distribution functions for the nine dSphs
                sorted by their absolute $M_V$ magnitude, from top to
                bottom and from left to right. The solid lines show
                the metallicity distribution convolved with the errors
                in metallicity. The dashed lines show the fitted
                gaussian distributions. The error bars in the upper
                right corner, or upper left in the case of DDO\,44,
                show the 1~$\sigma$ spread for the weighted mean
                metallicity we derive from our data. Note the
                different scaling of the individual y-axes.
               } 
  \label{sl_figure2}%
  \end{figure*}
%%%%%%%%%%%%%%%%%%%%%%%%%%%%%%%%%%%%%%%%%%%%%%%%%%%%%%%%%%%%%%%%%%%%%%%%%%%%%%%%%%%%
%
    \subsection{Photometric Metallicity Distribution Functions}
    \label{sec:mdfs}

    We show the photometric MDFs for the nine dSphs in
    Fig.~\ref{sl_figure2}. These are constructed using linear
    interpolation between Dartmouth isochrones (Dotter et
    al.~\cite{sl_dartmouth}) with a fixed age of 12.5~Gyr. We use
    Dartmouth isochrones, since they give the best simultaneous fit to
    the full stellar distribution within a CMD as demonstrated by
    e.g. Glatt et al.~(\cite{sl_glatta}, \cite{sl_glattb}). We chose
    the fixed age of 12.5~Gyr since the RGB in these dSphs may be
    assumed to consist of mainly old stars in an age range of about
    10~Gyr to 13~Gyr. The assumption of an old isochrone is also
    justified by the omnipresence of old stellar populations in all of
    the LG dSphs (Grebel \cite{sl_grebel01}; Grebel \& Gallagher
    \cite{sl_grebelgallagher}) and by the comparatively small number
    of luminous AGB stars above the TRGB.
    
    The choice of 12.5~Gyr is an assumption we are making in order to
    estimate the MDFs for each dwarf, while the choice of another age
    in the above range would not considerably affect our
    results. Indeed, the colors of the stars on the RGB are mostly
    affected by metallicity differences rather than age differences
    (see for example Caldwell et al.~\cite{sl_caldwell}; Harris,
    Harris \& Poole \cite{sl_harris99} (their Fig.~6); Frayn \&
    Gilmore \cite{sl_frayn} (their Fig.~2)). Thus the observed spread
    in the RGB color is likely caused by a metallicity spread rather
    than an age spread, justifying our choice of an constant age
    isochrone.

    The isochrone metallicities we used range from $-$2.50~dex to
    $-$0.50~dex with a step of 0.05~dex. The isochrone step we used is
    chosen such that it is smaller than the photometric
    errors. Representative photometric error bars are indicated with
    crosses in Fig.~\ref{sl_figure1}. To account for the influence of
    crowding and the photometric quality, we conducted artificial star
    tests. For that purpose, we used the utilities provided and as
    described in Dolphot.

    Given the lack of any spectroscopic information, this method of
    deriving the MDFs is in general fairly accurate as discussed in
    Frayn \& Gilmore\ (\cite{sl_frayn}). In practice, we interpolate
    between the two closest isochrones bracketing the color of a star,
    in order to find the metallicity of that star.

%%%%% FIGURE 3 - [Fe/H]_error versus F814W magnitude %%% One column figure%%%%%%
 \begin{figure}
  \centering
       \includegraphics[width=7.7cm,clip]{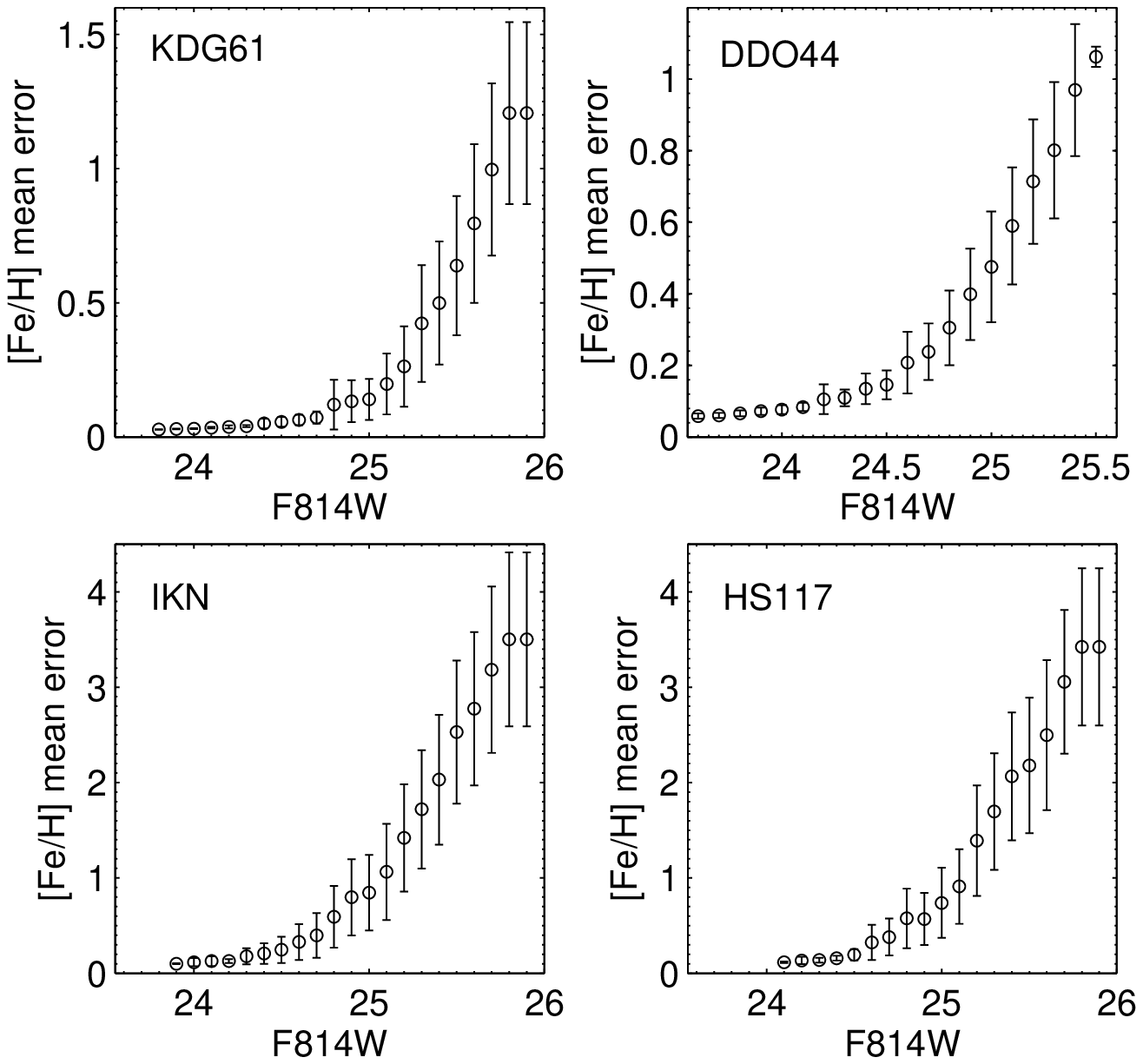}
       \caption{Mean error in metallicity versus the F814W-band
                magnitude. The circles indicate the mean error in
                metallicity in each magnitude bin, while the error
                bars indicate the standard deviation of the errors.}
  \label{sl_figure3}%
  \end{figure}
%%%%%%%%%%%%%%%%%%%%%%%%%%%%%%%%%%%%%%%%%%%%%%%%%%%%%%%%%%%%%%%%%%%%%%%%%%%%%%%%%%%%
%
    We only select stars within a box plausibly containing stars on
    the upper RGB to construct the galaxies' MDF. The bright magnitude
    limit of the box is chosen to exclude the stars brighter than the
    TRGB which belong mainly to the luminous AGB phase. The faint
    magnitude limit of the box is chosen to fulfil the requirement
    that the formal error in the derived [Fe/H] is less than 0.15~dex,
    or 0.2~dex in the case of IKN and HS\,117 when the photometric
    errors are taken into account. We employ a different selection
    criterion in the case of IKN and HS\,117 in order to have a
    significant number of stars in their sample as compared to the one
    of the remaining dSphs. The selection criterion based on the
    metallicity formal error depends on the depth of the
    observations. In our data sample we distinguish three categories,
    from now on referred to as depth categories. The first depth
    category contains \object{KDG\,61}, \object{KDG\,64},
    \object{DDO\,71}, F6D1 and F12D1. The second depth category
    contains \object{DDO\,44} and \object{DDO\,78}. The third depth
    category contains \object{IKN} and HS\,117. Each depth category
    contains those dSphs that belong to the same Program~ID and thus
    have the same filters and roughly the same exposure times, as
    listed in Table~\ref{table1}.

    In order to estimate the faint magnitude limit for each dSph's RGB
    box as a function of the error in [Fe/H], we proceed as
    follows. We extend the faint limit of the bounding box to a
    magnitude limit of 26 in F814W for all the dSphs. We compute the
    [Fe/H] for all the stars within each dSph's box, as well as the
    corresponding errors in metallicity. We show the derived mean
    errors in metallicity versus the F814W-band magnitude in
    Fig.~\ref{sl_figure3} for KDG\,61, DDO\,44, IKN and HS\,117, which
    are chosen here as representative  examples of the three depth
    categories. In the case of IKN and HS\,117, which belong to the
    third depth category, we show the corresponding plots for both
    since the requirement of 0.20~dex leads to slightly different
    faint limits of the RGB box. Based on these plots and on the
    metallicity requirements, we choose 25 and 24.5~mag as faint limit
    for the first and second depth category, while in the case of IKN
    and HS\,117 we choose 24.4 and 24.6~mag, respectively. The choice
    of these limits corresponds to an error in color of less than
    0.02~mag for the first depth category and less than 0.07~mag for
    the remaining two depth categories. We note that the difference in
    the error in color is due to the different exposure times for each
    dSph data set, which are listed in the column (6) of
    Table~\ref{table1}. The RGB boxes used in each dSph are drawn in
    Fig.~\ref{sl_figure1}.

%%%%% FIGURE 4 - RGB isochrone - [Fe/H] error %% Figure with a new BoundingBox %%%%%%%
    \begin{figure}
    \centering
    \includegraphics[width=8.5cm,clip]{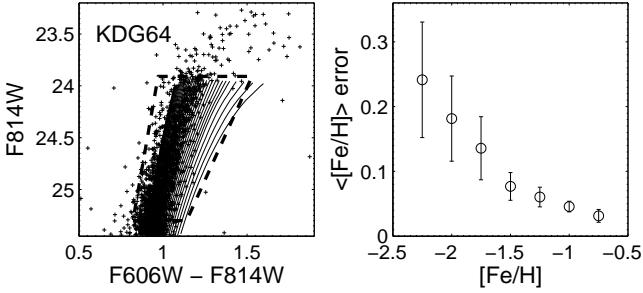}
       \caption{\textit{Left panel}: Color-magnitude diagram for
                KDG\,64 zoomed in the RGB part to show the stars
                selected in the box, shown with the dashed line, for
                which we compute their photometric metallicities. In
                the same figure we overplot with solid lines a subset
                of the isochrones used for the interpolation
                method, with metallicities ranging from
                  $-$2.50~dex to $-$0.80~dex.
                \textit{Right panel}: The error in metallicity versus
                the [Fe/H] as derived with the use of Monte Carlo
                simulations. The circles indicate the mean metallicity in
                each metallicity bin, while the error bars indicate
                the standard deviation of the mean.}
    \label{sl_figure4}%
    \end{figure}
%%%%%%%%%%%%%%%%%%%%%%%%%%%%%%%%%%%%%%%%%%%%%%%%%%%%%%%%%%%%%%%%%%%%%%%%%%%%%%%%%%%%
%
    In Fig.~\ref{sl_figure4}, left panel, we plot the RGB box used in
    the case of the dwarf KDG\,64 as well as a subset of the
    isochrones used for the interpolation method, here ranging from
    $-$2.50~dex to $-$0.80~dex. The step remains 0.05~dex. The grid of
    the theoretical isochrones we use is fine enough that the spacing
    between them is kept nearly vertical. We correct the magnitudes
    and colors of the theoretical isochrones for foreground Galactic
    extinction and for the distance modulus of each dSph. The
    $A_{F814W}$ and $A_{F606W}$ (or $A_{F475W}$ in the case of DDO\,44
    and DDO\,78) that we calculate and the true distance moduli are
    listed in Table~\ref{table2}, columns (6) and (7) for the
    extinction and (8) for the distance moduli. The I-band TRGB values
    shown in column (5) of the same table were used to compute the
    F814W-band TRGB values, as explained already in
    Sec.~\ref{sec:observations}.

    We note in Fig.~\ref{sl_figure4} the presence of stars within the
    RGB box bluewards of the most metal-poor isochrone available from
    the Dartmouth set of isochrones. These stars are not used to
    construct the MDF. The existence of such stars may indicate the
    presence of more metal-poor RGB stars or of old AGB stars, with
    ages typically greater or equal to 10~Gyr. Such old AGB stars that
    have the same luminosity as RGB stars were also noted, for
    example, by Harris, Harris \& Poole (\cite{sl_harris99}) while
    constructing the MDF for stars in a halo field of the giant
    elliptical NGC\,5128. It is expected that at most 22~\% of the
    stars in the RGB selection box, and within 1~mag below the TRGB,
    are actually old AGB stars (Durrell, Harris \& Pritchet
    \cite{sl_durrell01}; Martinez-Delgado \& Aparicio
    \cite{sl_martinez97}; and references therein). In order to
    quantify the effect of the presence of such stars, we construct
    the MDF of KDG\,61 using Padova isochrones (Girardi et
    al.~\cite{sl_girardi08}; Marigo et al.~\cite{sl_marigo08}), which
    also include the AGB phase. We run the interpolation code once
    using isochrones that only include the RGB phase and once with
    isochrones that only include the AGB phase, for a constant age of
    12.5~Gyr and a range in metallicities from [Fe/H]\,$=\,-$2.36~dex
    (or Z\,$=$\,0.0001) to [Fe/H]\,$=\,-$0.54~dex (or Z\,$=$\,0.006),
    with a step of 0.1~dex in [Fe/H]. The derived mean values differ
    only by 0.04~dex in [Fe/H] with a mean of
    $\langle$[Fe/H]$\rangle\,=\,-$1.24~dex for the MDF constructed
    using isochrones that include only the RGB phase. The MDF of the
    stars that were fit using isochrones that include only the AGB
    phase becomes more metal-rich. Furthermore, the derived 1~$\sigma$
    spreads in [Fe/H] have comparable values of 0.26~dex when we
    include only the RGB phase and of 0.27~dex when we include only
    the AGB phase. In addition, if we randomly assign 22~\% of the
    stars within the RGB box with metallicities as derived using only
    the AGB phase, while the remaining 78~\% of the stars with
    metallicities as derived using only the RGB phase, then the
    resulting MDF has a mean of $\langle$[Fe/H]$\rangle\,=\,-$1.24~dex
    with 1~$\sigma$ spread in [Fe/H] of 0.29~dex. This mean
    metallicity is comparable to the one we compute when we use only
    the RGB phase to derive the metallicities. The shape of the MDFs
    in all these cases does not change significantly. Thus, we can
    safely conclude that the presence of these contaminating old AGB
    stars within the RGB box does not affect the derived MDFs'
    properties significantly.

%%%%% TABLE 3 %%%%%%%%%%%%%%%%%%%%%%%%%%%%%%%%%%%%%%%%%%%%%%%%%%%%%%%%%%%%%%%%%%%%%%
\begin{table}
\begin{minipage}[t]{\columnwidth}
\caption[]{Derived Properties. }
\label{table3} 
\centering
\renewcommand{\footnoterule}{}  % to avoid a line before footnotes
\begin{tabular}{l c c c c c}     % 6 columns  

\hline\hline
  Galaxy    &$\langle$[Fe/H]$\rangle\pm\sigma$  &$\langle$[Fe/H]$\rangle_{w}\pm\sigma$   &$K-S$\footnote{The probabilities indicate whether the populations under conside\-ration are from the same distribution.}              
                                                                                                                             &$f_{AGB}$    \\
            &(dex)                              &(dex)                                   &(\%)                               &            \\
   (1)      &(2)                                &(3)                                     &(4)                                &(5)         \\
                                                                                                                           
\hline                                                                                                                      
                                                                                                                            
  KDG\,61   &$-1.65\pm0.28$                     &$-1.49\pm0.26$                          &$16$                               &$0.07$       \\
                                                                                                                              
  KDG\,64   &$-1.72\pm0.30$                     &$-1.57\pm0.23$                          &$12$                               &$0.09$       \\
                                                                                                                                  
  DDO\,71   &$-1.64\pm0.29$                     &$-1.56\pm0.24$                          &$0$                                &$0.09$       \\

  F12D1     &$-1.56\pm0.27$                     &$-1.43\pm0.34$                          &$8$                                &$0.07$       \\

  DDO\,78   &$-1.51\pm0.35$                     &$-1.36\pm0.20$                          &$0$                                &$0.09$       \\
                                                                                                                                      
  DDO\,44   &$-1.77\pm0.29$                     &$-1.67\pm0.19$                          &$0$                                &$0.11$       \\
                                                                                                                                    
  IKN	    &$-1.38\pm0.37$                     &$-1.08\pm0.16$                          &...                                &$0.08$       \\
                                                                                                                                    
  F6D1	    &$-1.63\pm0.30$                     &$-1.48\pm0.43$                          &$0.036$                            &$0.03$       \\

  HS\,117   &$-1.65\pm0.32$                     &$-1.41\pm0.20$                          &$55$                               &$0.14$       \\
                                                                                                   
\hline
 
\end{tabular}
\end{minipage}
\end{table}
%%%%%%%%%%%%%%%%%%%%%%%%%%%%%%%%%%%%%%%%%%%%%%%%%%%%%%%%%%%%%%%%%%%%%%%%%%%%%%%%%%%%
%
    In Fig.~\ref{sl_figure2} we overplot the metallicity distribution
    convolved with the errors in metallicity (solid line). Also shown
    in Fig.~\ref{sl_figure2} (dashed lines) are fits of Gaussian
    distributions with the observed mean and dispersion. For each dSph
    we compute the mean metallicity, $\langle$[Fe/H]$\rangle$, as well
    as the error-weighted mean metallicity,
    $\langle$[Fe/H]$\rangle_w$, along with the corresponding intrinsic
    1~$\sigma$ dispersions. We show them in Table~\ref{table3},
    columns (2) and (3), while the error bars in Fig.~\ref{sl_figure2}
    indicate the 1~$\sigma$ dispersion of the error-weighted mean
    metallicities. The errors in metallicity are computed from a set
    of Monte Carlo simulations, in which each star is varied by its
    photometric uncertainties (both in color and magnitude, as given
    by the Dolphot output) and re-fit using the identical isochrone
    interpolation as described above. The 1~$\sigma$ scatter of the
    output random realizations was then adopted as the metallicity
    error for each star. In the right panel of Fig.~\ref{sl_figure4}
    we show the errors in metallicity computed as described above
    versus the metallicities derived for all the stars within the RGB
    box, here for KDG\,64 as an example. The error in metallicity
    increases towards the metal-poor part, which is due to the spacing
    between the isochrones that becomes narrower towards the
    metal-poor part.

%%%%% FIGURE 5 - MDFs YOUNG-OLD %%% One column figure %%%%%%%%%%%%%%%
 \begin{figure}
  \centering
       \includegraphics[width=7.8cm,clip]{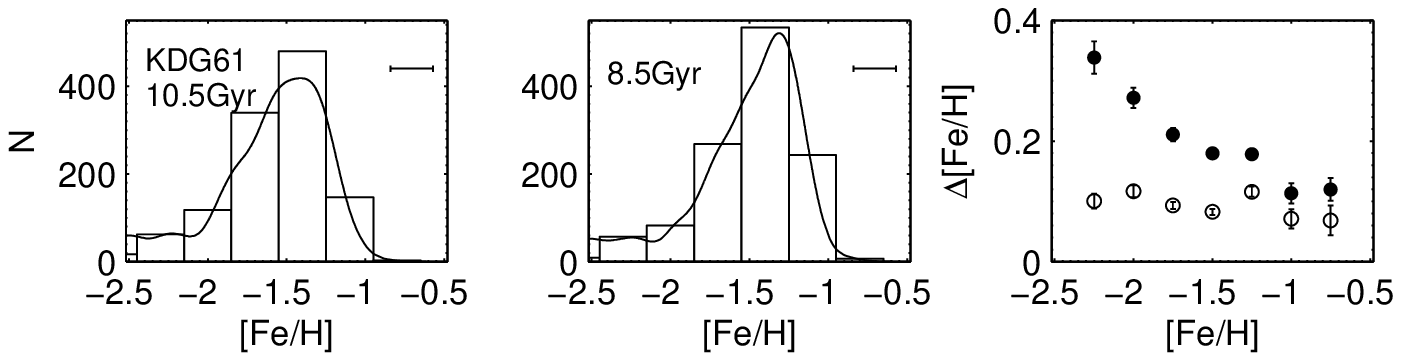}
       \includegraphics[width=7.8cm,clip]{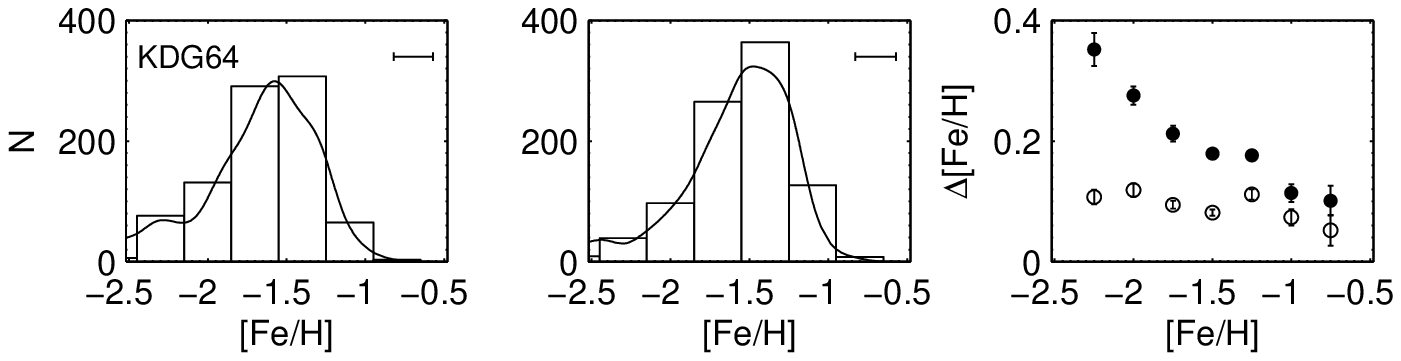}
       \includegraphics[width=7.8cm,clip]{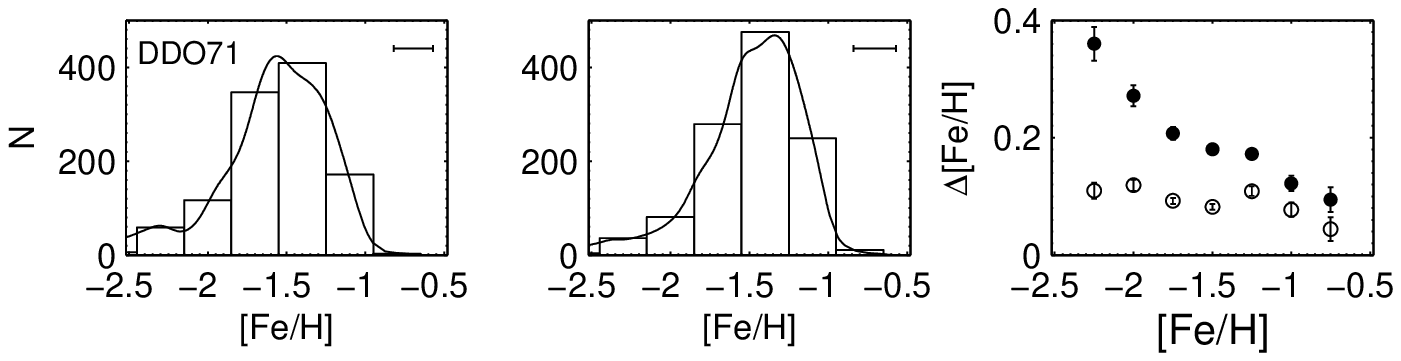}
       \includegraphics[width=7.8cm,clip]{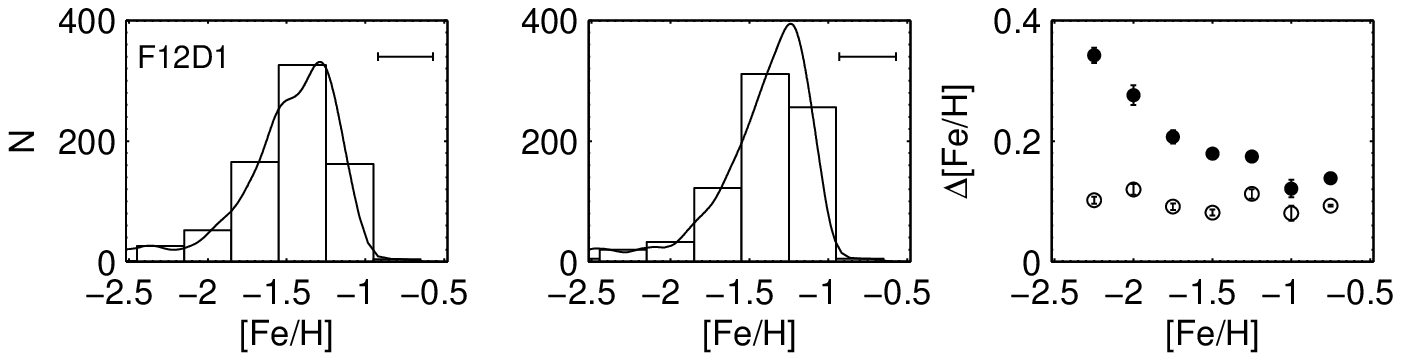}
       \includegraphics[width=7.8cm,clip]{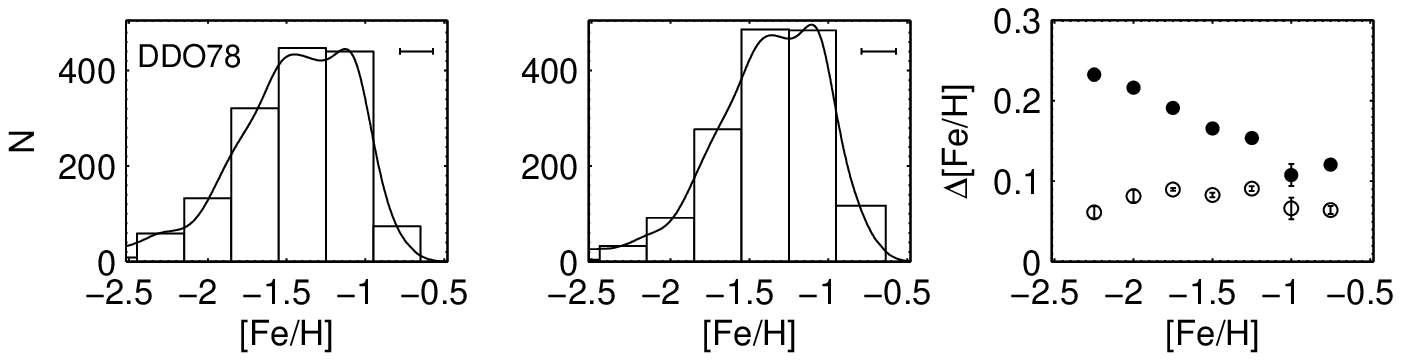}
       \includegraphics[width=7.8cm,clip]{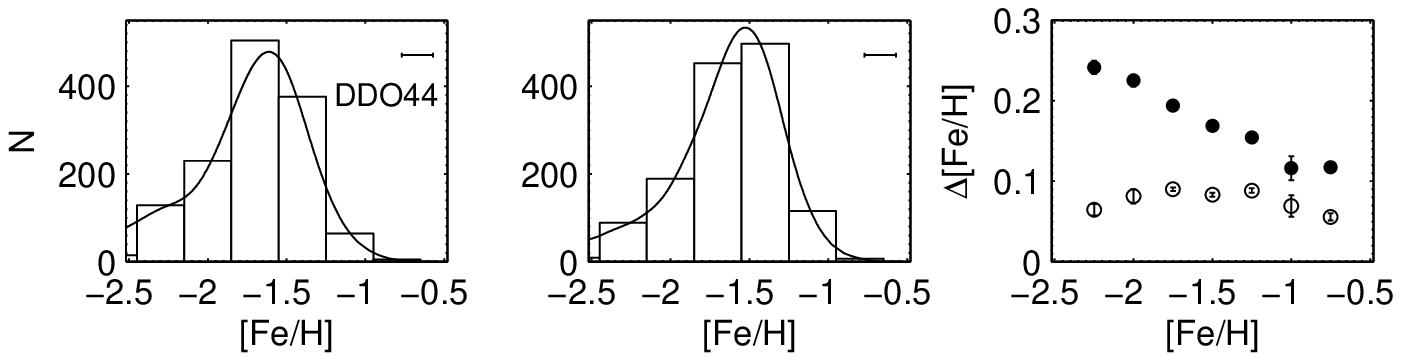}
       \includegraphics[width=7.8cm,clip]{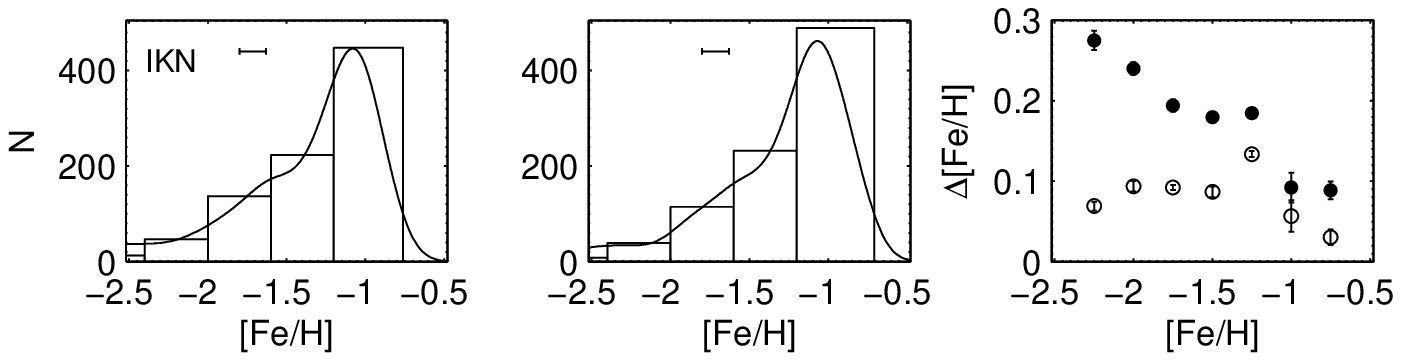}
       \includegraphics[width=7.8cm,clip]{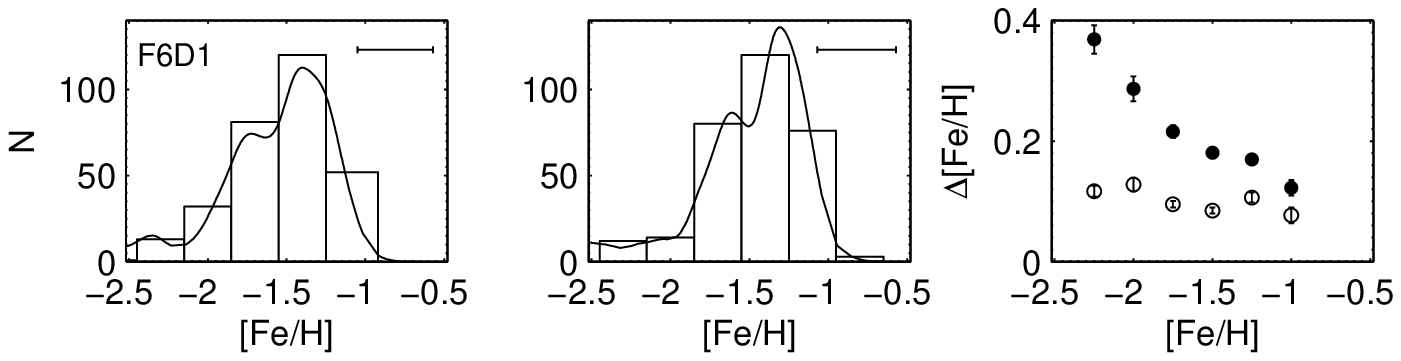}
       \includegraphics[width=7.8cm,clip]{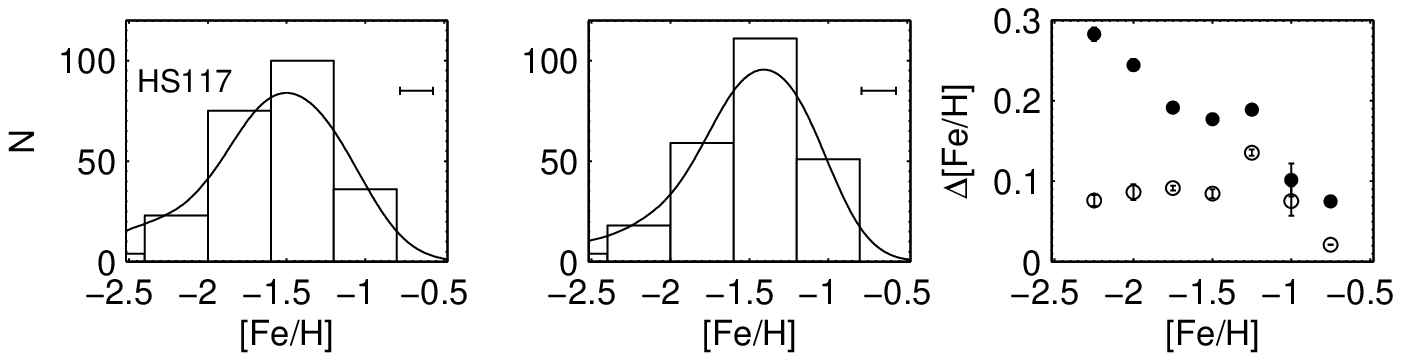}
       \caption{Metallicity distribution functions for the nine dSphs
                computed in the same way using isochrones with two
                different ages: \textit{left panels} for 
                a constant age of 10.5~Gyr and \textit{middle
                  panels} for a constant age of
                  8.5~Gyr. The solid lines show the metallicity  
                distribution convolved with the errors in
                metallicity. The error bars correspond to the 
                1~$\sigma$ spread for the weighted mean metallicity we
                derive from our data.
                \textit{Right panels:} Star-by-star difference of the
                derived metallicities, $\Delta$[Fe/H], using the
                10.5~Gyr isochrones minus the 12.5~Gyr
                  isochrones versus the [Fe/H] of the 10.5~Gyr
                  isochrones, indicated with the open circles. The
                  star-by-star differences of the 8.5~Gyr isochrones
                  minus the 12.5~Gyr isochrones are indicated with the
                  dots.
                }
  \label{sl_figure5}%
  \end{figure}
%%%%%%%%%%%%%%%%%%%%%%%%%%%%%%%%%%%%%%%%%%%%%%%%%%%%%%%%%%%%%%%%%%%%%%%%%%%%%%%%%%%%
%
%%%%% TABLE 4%%%%%%%%%%%%%%%%%%%%%%%%%%%%%%%%%%%%%%%%%%%%%%%%%%%%%%%%%%%%%%%%%%
\begin{table}
\begin{minipage}[t]{\columnwidth}
\caption[]{Error-weighted mean metallicities for the 10.5~Gyr and 8.5~Gyr isochrones MDFs.}
\label{table4} 
\centering
\begin{tabular}{l c c }     % 3 columns  

\hline\hline
  Galaxy     &$\langle$[Fe/H]$\rangle_{w,10.5}\pm\sigma$  &$\langle$[Fe/H]$\rangle_{w,8.5}\pm\sigma$  \\
             &(dex)                                 &(dex)                                  \\
   (1)       &(2)                                   &(3)                                    \\
                            
\hline                       
                             
  KDG\,61    &$-1.37\pm0.27$                        &$-1.32\pm0.27$ \\           
                                                       
  KDG\,64    &$-1.45\pm0.25$                        &$-1.39\pm0.26$ \\           
                                                           
  DDO\,71    &$-1.41\pm0.25$                        &$-1.34\pm0.27$ \\             
                                                                       
  F12D1      &$-1.31\pm0.35$                        &$-1.25\pm0.36$ \\           
                                                                      
  DDO\,78    &$-1.28\pm0.21$                        &$-1.24\pm0.22$ \\           
                                                                
  DDO\,44    &$-1.60\pm0.20$                        &$-1.53\pm0.20$ \\           
                                                             
  IKN	     &$-1.03\pm0.17$                        &$-0.98\pm0.17$ \\             
                                                             
  F6D1	     &$-1.37\pm0.48$                        &$-1.31\pm0.50$ \\
                                                                     
  HS\,117    &$-1.31\pm0.21$                        &$-1.27\pm0.22$ \\           
                                                                                                   
\hline
 
\end{tabular}
\end{minipage}
\end{table}
%%%%%%%%%%%%%%%%%%%%%%%%%%%%%%%%%%%%%%%%%%%%%%%%%%%%%%%%%%%%%%%%%%%%%%%%%%%%%%%%%%%%
    In order to further quantify the effect of the assumption of the
    constant age on the MDFs, we apply again the same analysis using
    two different constant ages for the isochrones in the
    interpolation method. The first constant age for the isochrones is
    10.5~Gyr, while the second constant age is 8.5~Gyr. We repeat the
    isochrone interpolation with the bounding boxes and the
    metallicity ranges being kept the same in all cases. We show the
    results for the MDFs of all the dSphs in Fig.~\ref{sl_figure5},
    where the MDFs for the 10.5~Gyr isochrones are shown in the left
    panels and the ones for the 8.5~Gyr isochrones in the middle
    panels. The derived error-weighted mean metallicities
    $\langle$[Fe/H]$\rangle_{w,10.5}$ and
    $\langle$[Fe/H]$\rangle_{w,8.5}$, for the 10.5~Gyr and 8.5~Gyr
    isochrones, respectively, are shown in Table~\ref{table4}, along
    with their corresponding dispersions.

    In addition, the star-by-star difference in [Fe/H] as derived
    using the 10.5~Gyr and 8.5~Gyr isochrones is shown in
    Fig.~\ref{sl_figure5}, right panels. The maximum difference in the
    derived [Fe/H] using the 10.5~Gyr isochrones minus the isochrones
    with a constant age of 12.5~Gyr is less than 0.20~dex in all the
    cases, while the maximum difference in the derived [Fe/H] using
    the 8.5~Gyr isochrones minus the isochrones with a constant age of
    12.5~Gyr is less than 0.40~dex in all the cases. Finally, the
    overall shape of the MDFs as derived for the 10.5~Gyr and the
    8.5~Gyr isochrones does not change significantly.

    \subsection{Population gradients}
    \label{sec:gradients}
 
    In order to examine the presence or absence of population
    gradients in our dSph sample, we construct the cumulative
    histograms of the stars in each dSph selected in two metallicity
    ranges, defined as above and below the respective
    $\langle$[Fe/H]$\rangle_{w}$. Since the dSphs can be considered as
    being elliptical in projection to first order, we define in the
    following the elliptical radius $r$ as   
    \begin{equation} 
        r = \sqrt {x^2 + \frac{y^2} {(1-\epsilon)^2}},
    \end{equation}
    where $x$ and $y$ are the distance along the major and minor axis,
    and $\epsilon$ is the ellipticity. The major and minor axes are
%
%%%%% FIGURE 6 - CONTOURS %%% Two columns figure %%%%%%%%%%%%%%%
 \begin{figure*}
  \centering
       \includegraphics[width=6cm,clip]{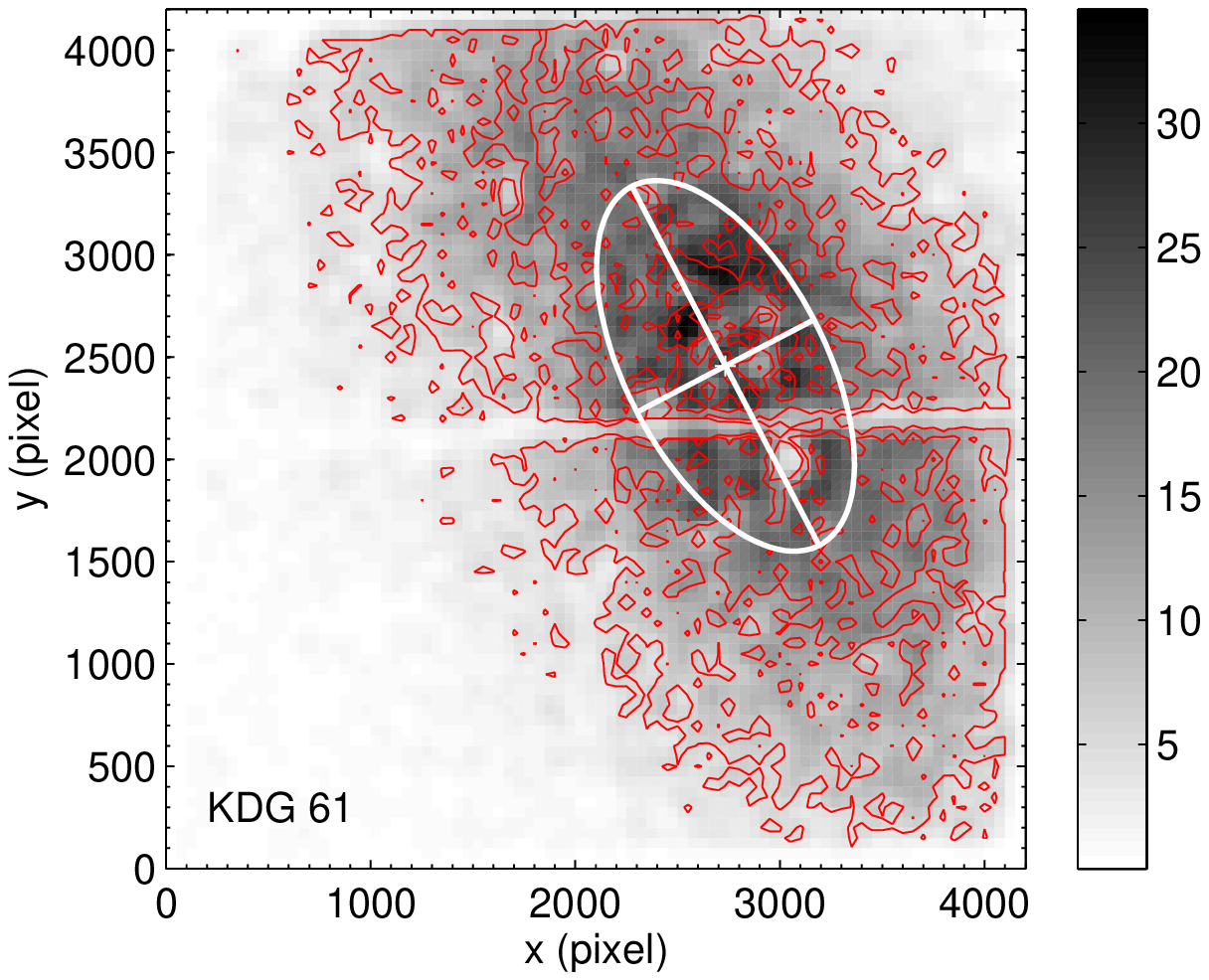}
       \includegraphics[width=6cm,clip]{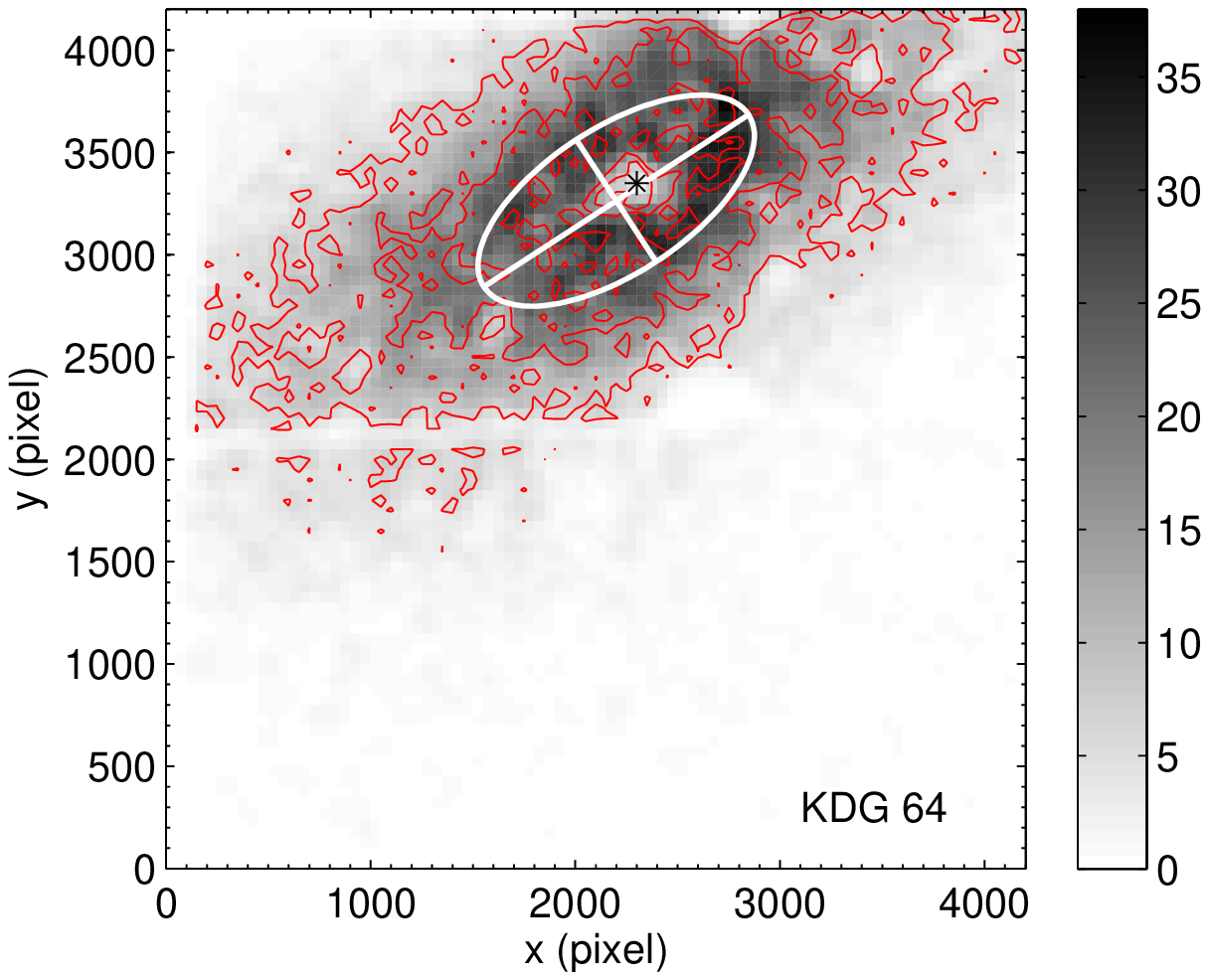}
       \includegraphics[width=6cm,clip]{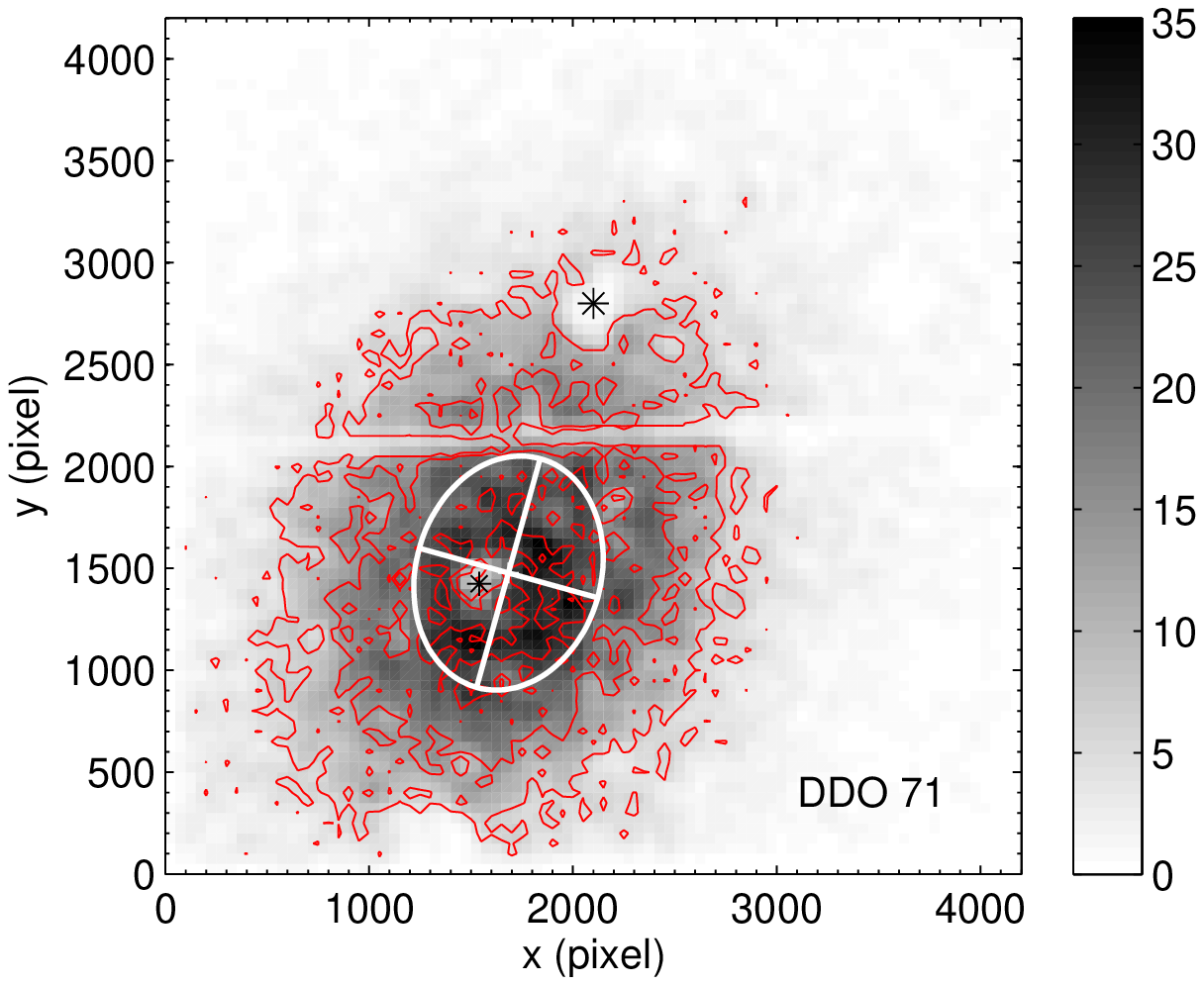}
       \includegraphics[width=6cm,clip]{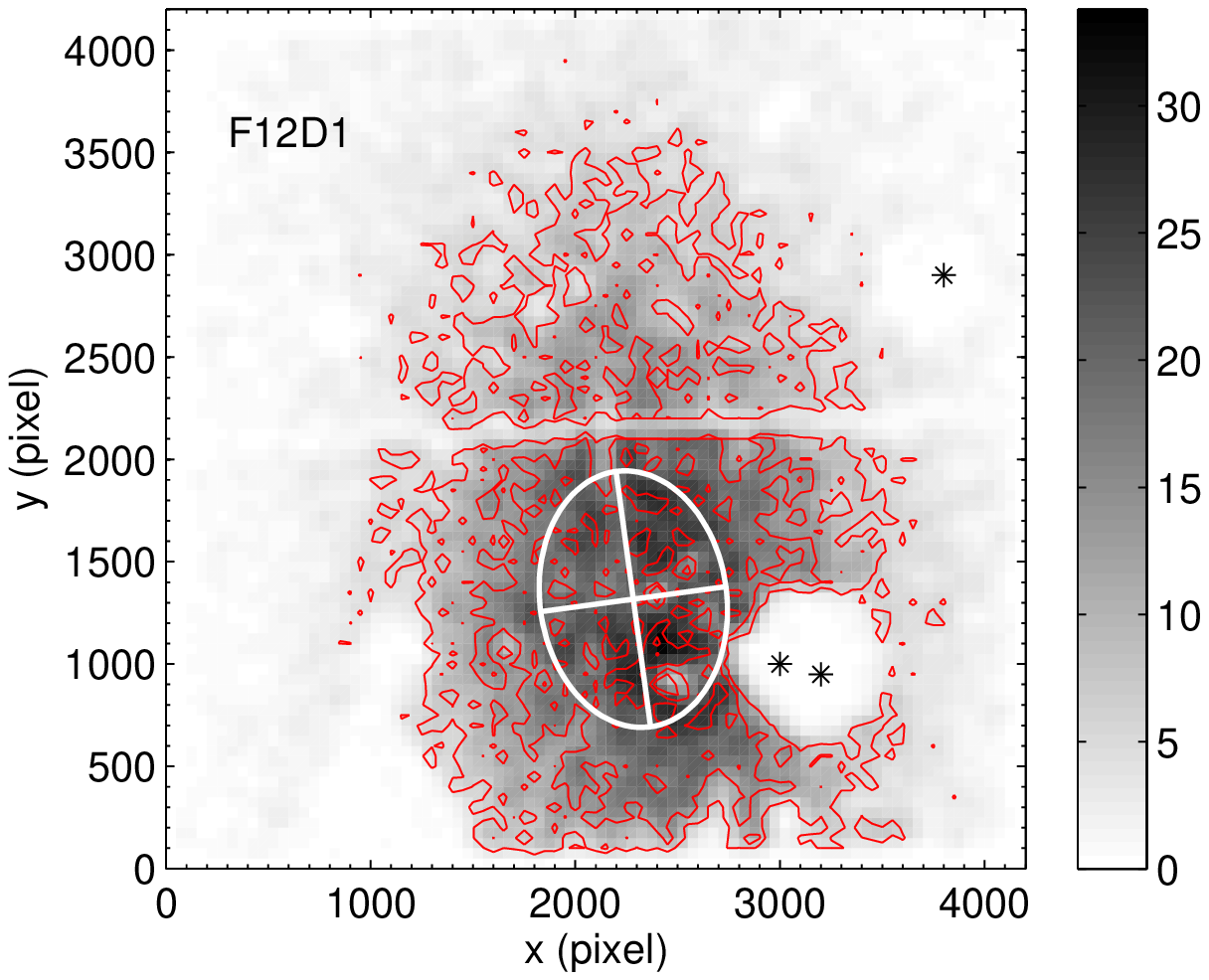}
       \includegraphics[width=6cm,clip]{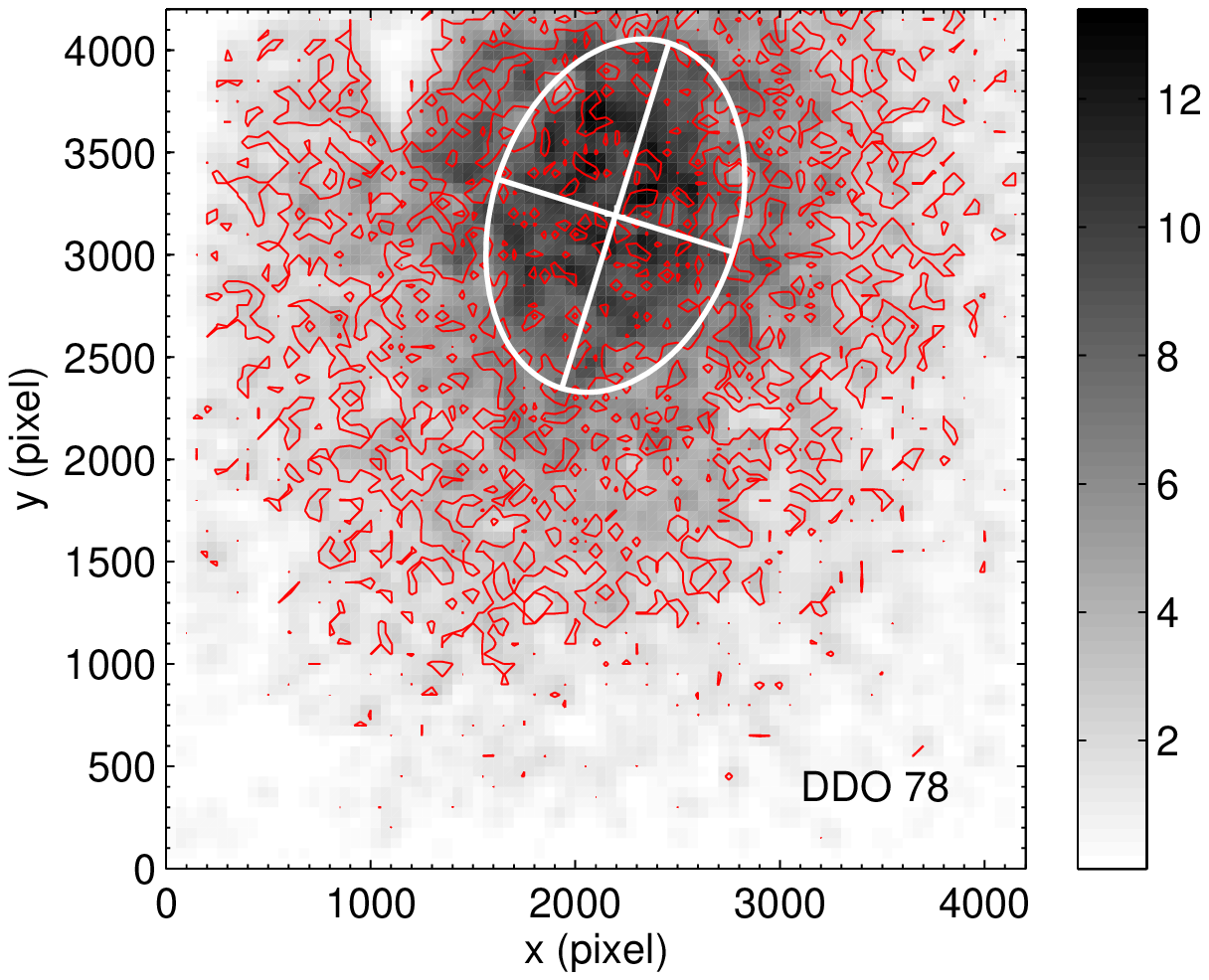}
       \includegraphics[width=6cm,clip]{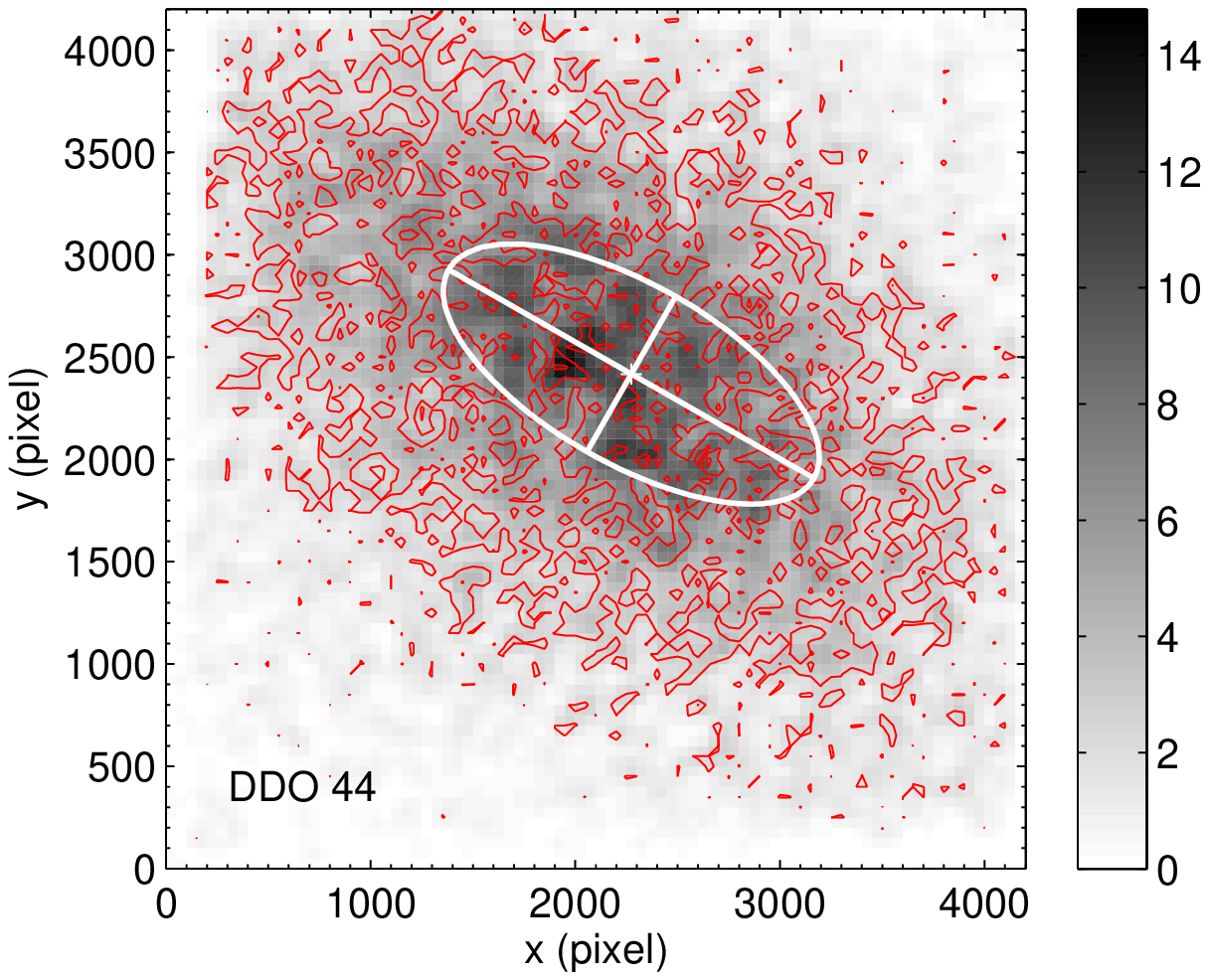}
       \includegraphics[width=6cm,clip]{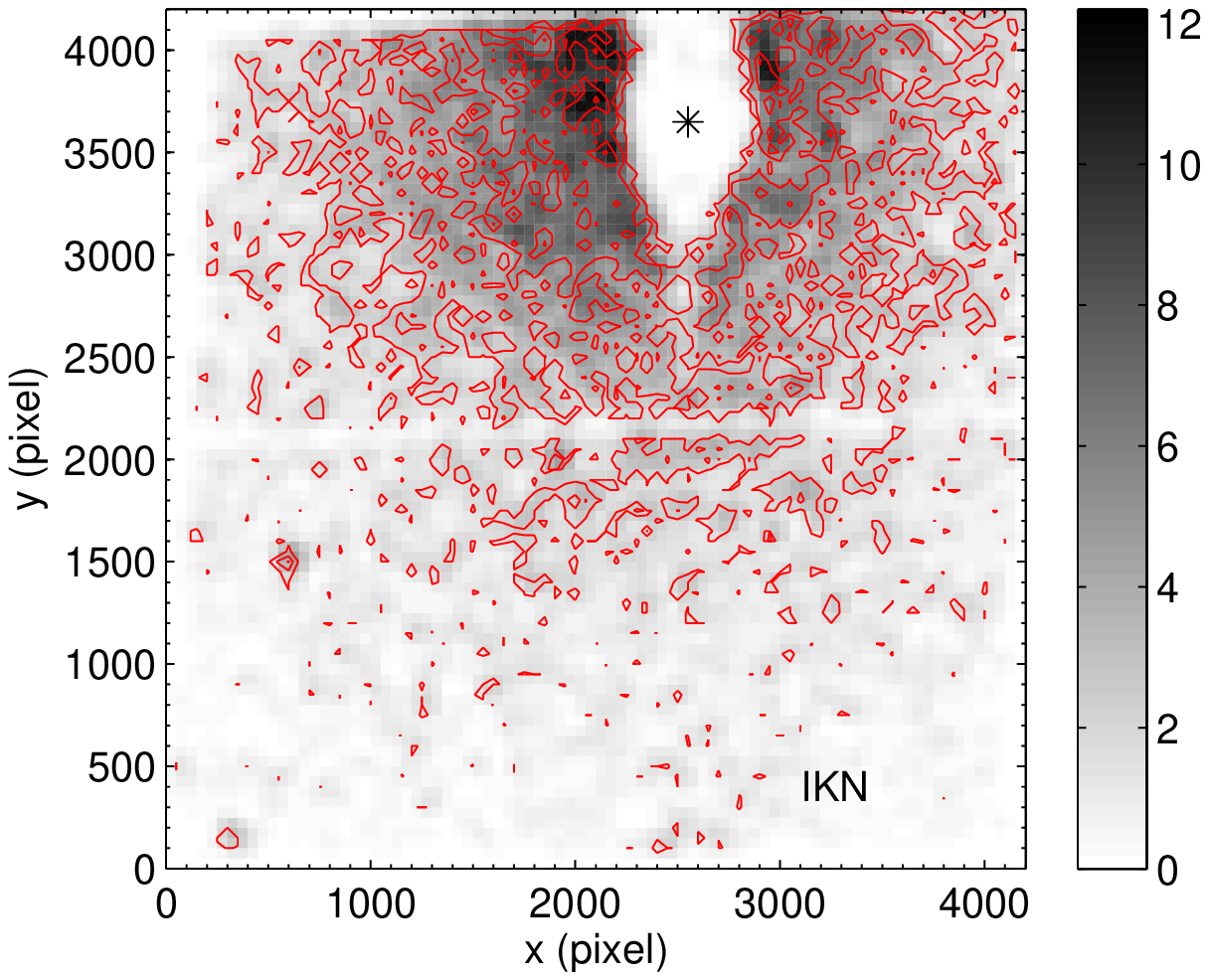}
       \includegraphics[width=6cm,clip]{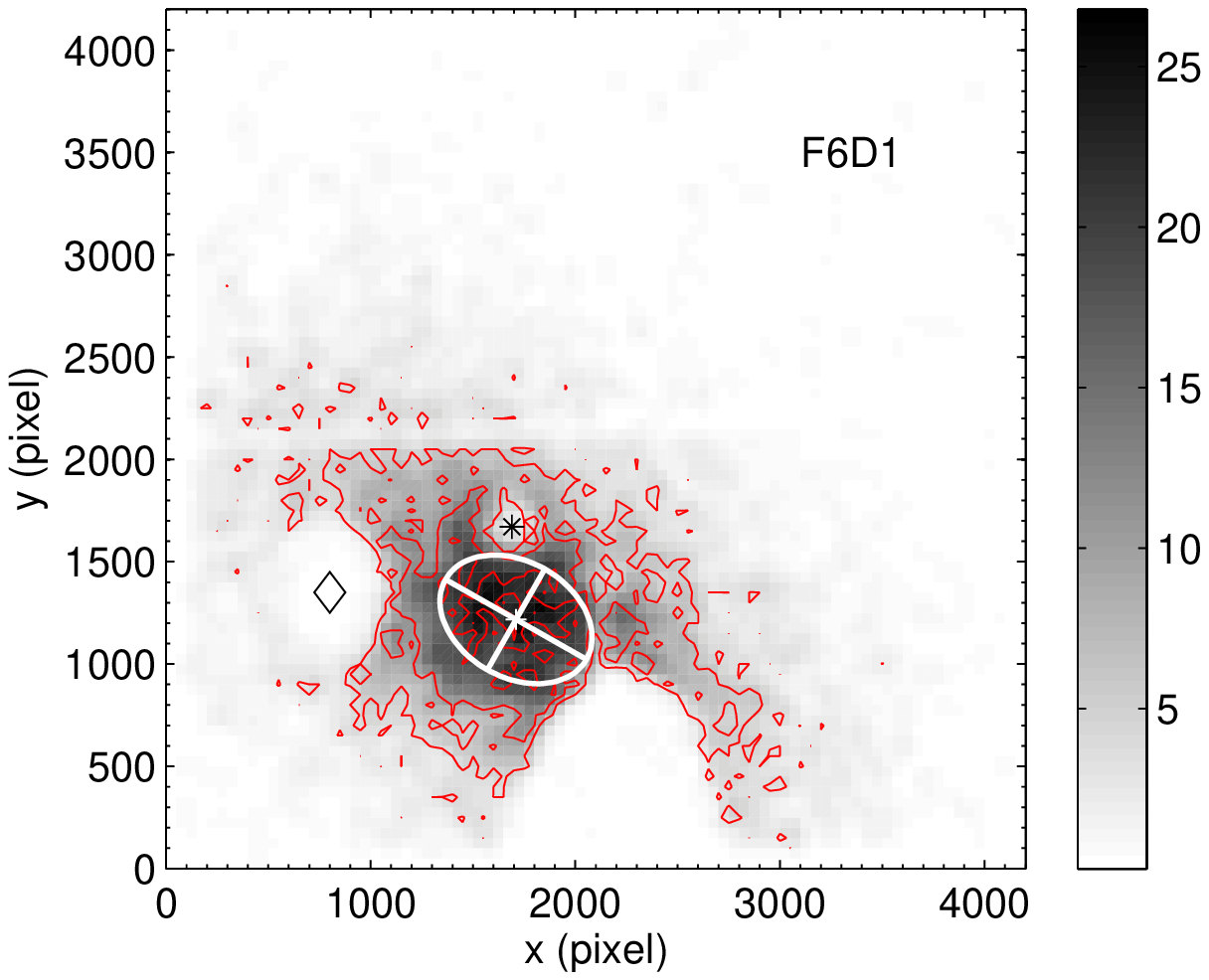}
       \includegraphics[width=6cm,clip]{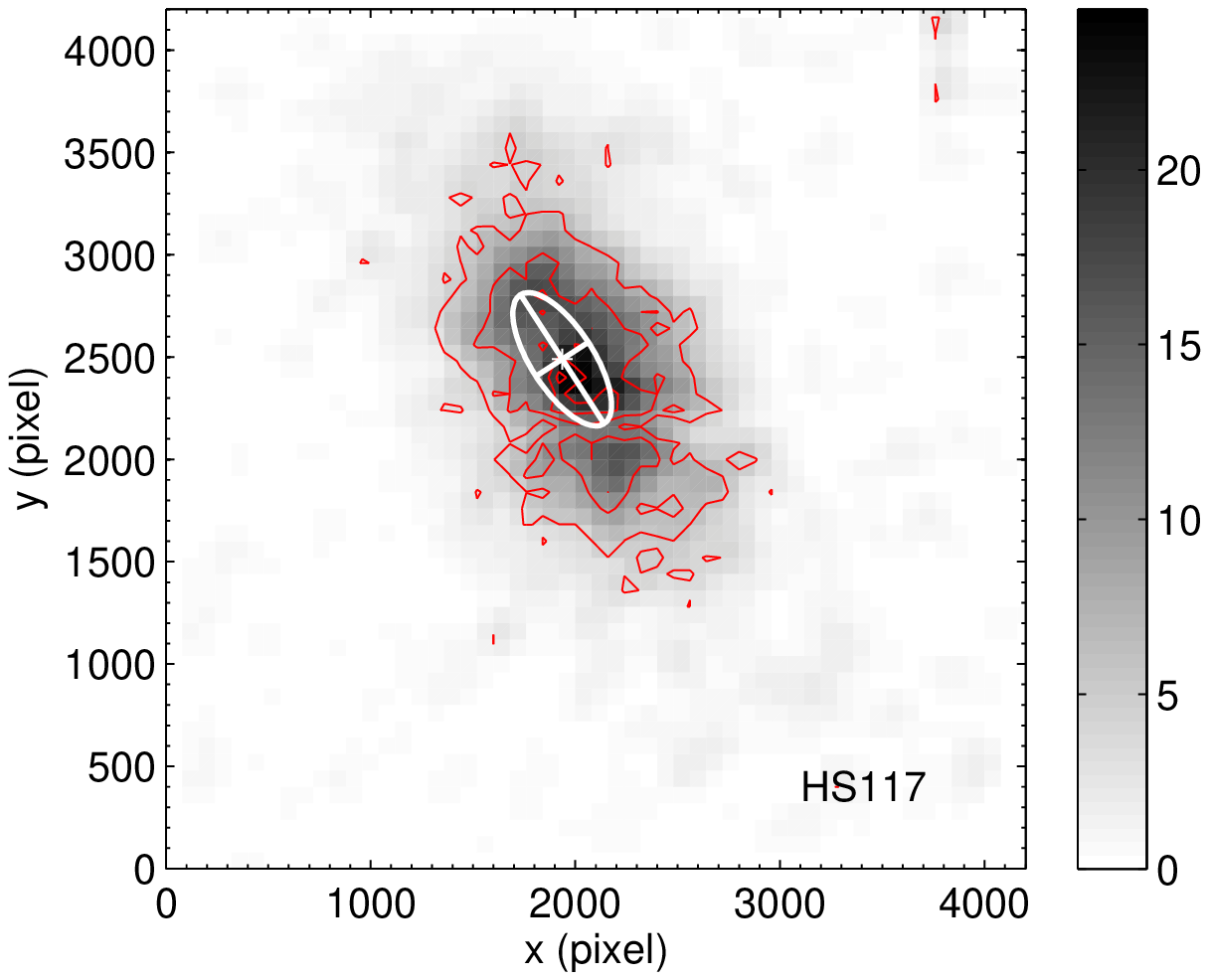}
       \caption{Contour plots, shown in red, for the nine dSphs which
                are overlaid on top of density maps. The elliptical
                shape chosen for each dSph, with the exception of IKN,
                is shown with the white ellipse. The star symbols in
                some plots correspond to bright foreground stars,
                while in the case of KDG\,64 corresponds to a
                background galaxy. The diamond symbol in the case of 
                F6D1 corresponds to an extended background galaxy. The
                unit of the colorbars is number of stars per
                (50~pixels)$^2$.}
  \label{sl_figure6}%
  \end{figure*}
%%%%%%%%%%%%%%%%%%%%%%%%%%%%%%%%%%%%%%%%%%%%%%%%%%%%%%%%%%%%%%%%%%%%%%%%%%%%%%%%%%%%
    computed by fitting an ellipse to contours of the number counts of
    all stars above the 1~$\sigma$ level. This ellipse is shown in
    white in Fig.~\ref{sl_figure6} and represents the elliptical shape
    that was chosen for each dSph. In the same Fig.~\ref{sl_figure6}
    we show the contours above the 0.5~$\sigma$ to 2~$\sigma$ level,
    which are overlaid on top of density maps. In the study of
    population gradients we exclude the IKN dSph since the field of
    view does not cover the whole extent of the galaxy and furthermore
    is contaminated by a bright foreground star. In addition, we do
    not show the elliptical shape for IKN.

%%%%% FIGURE 7 - Cumulatives1 %%%%% A figure as large as the width of the column %%%
   \begin{figure*}
   \centering
   \includegraphics[width=6cm,clip]{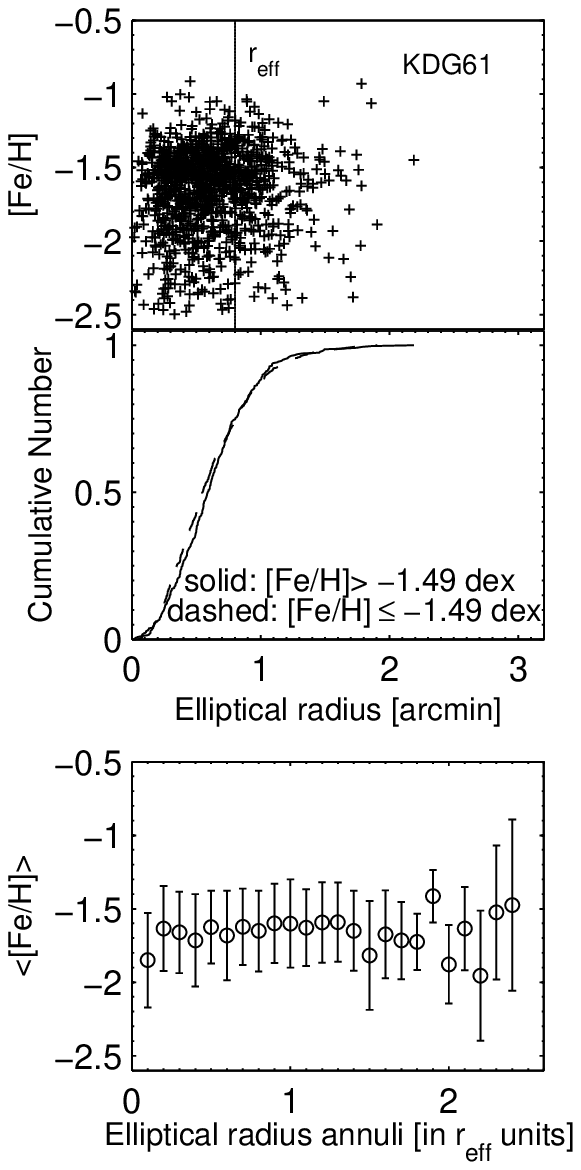}
   \includegraphics[width=6cm,clip]{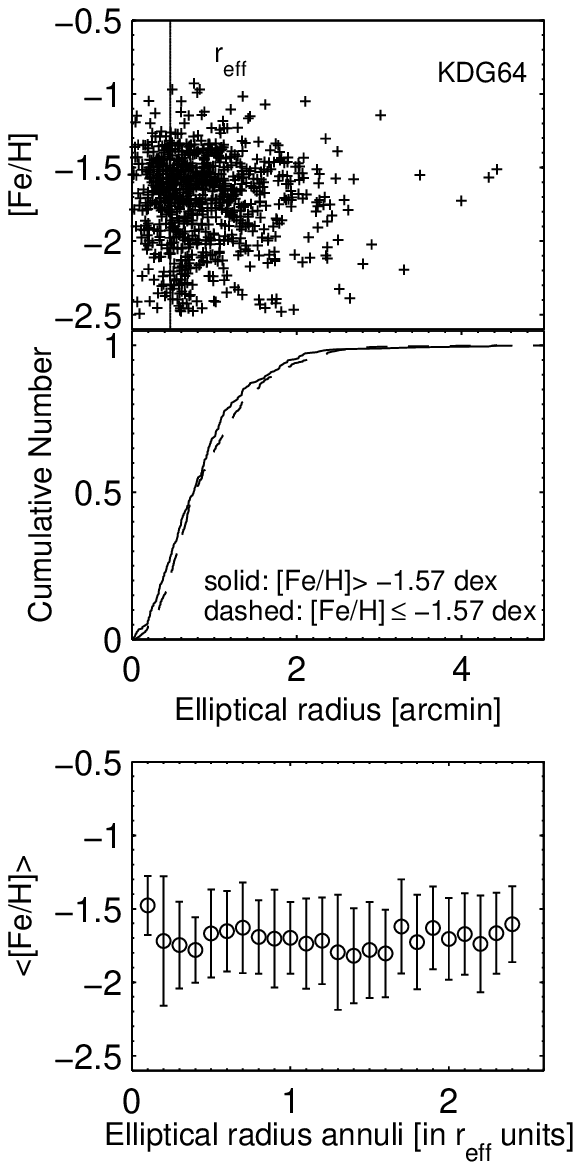}
   \includegraphics[width=6cm,clip]{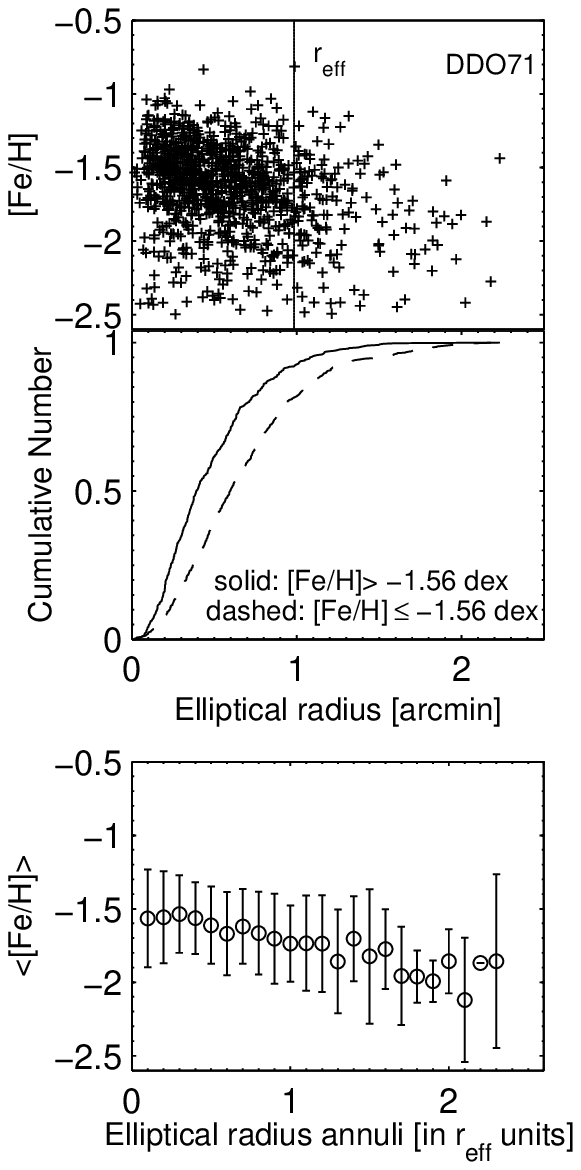}
   \includegraphics[width=6cm,clip]{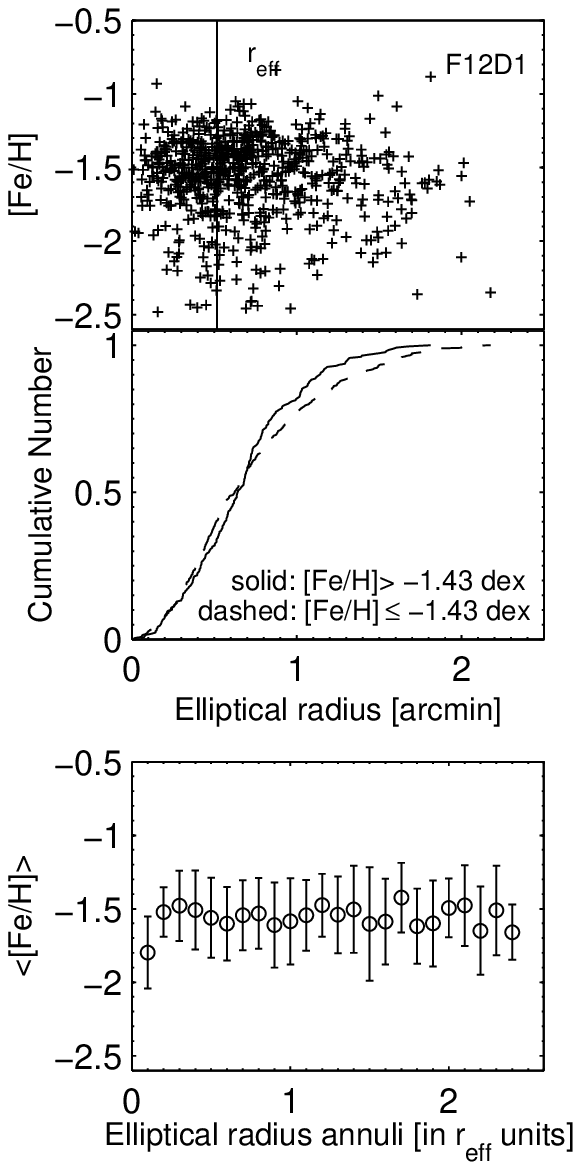}
   \includegraphics[width=6cm,clip]{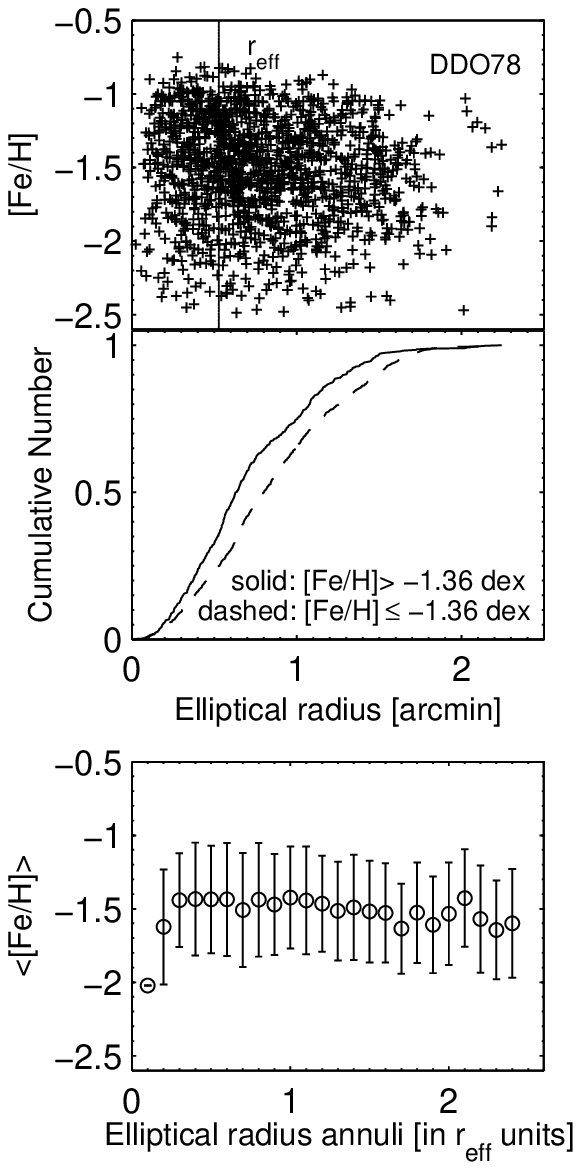}
   \includegraphics[width=6cm,clip]{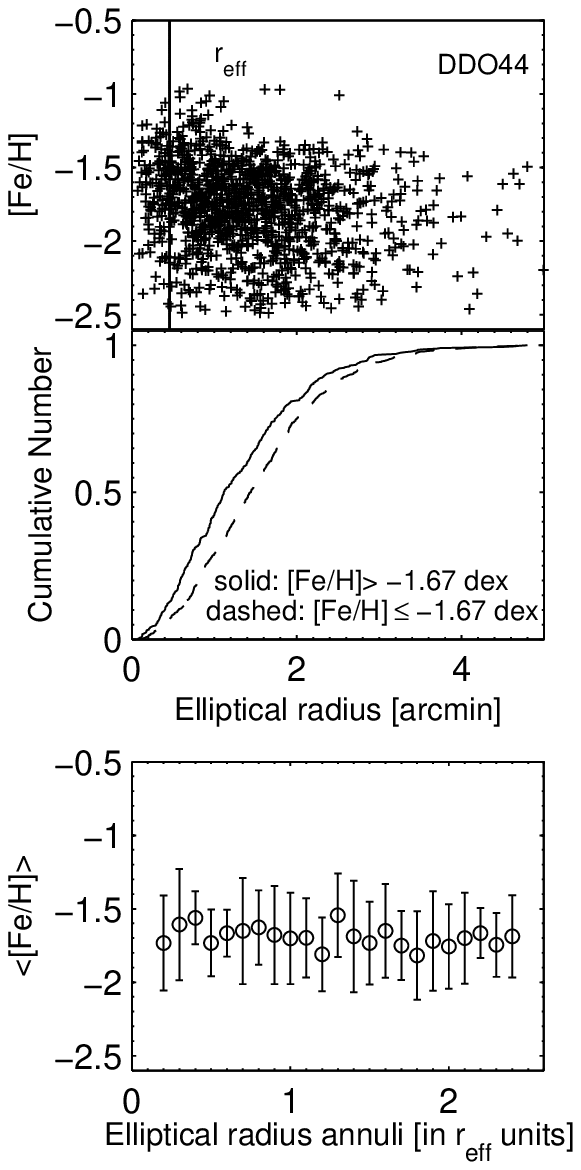}
      \caption{In each panel from left to right and from top to bottom
               we show the radial metallicity distributions (top
               panels), their cumulative distributions (middle
               panels), and the radial mean metallicity profile
               (bottom panels). The cumulative histogram is for stars
               selected in metallicity above and below the weighted
               mean value $\langle$[Fe/H]$\rangle_{w}$ listed in
               Table~\ref{table3}.}
   \label{sl_figure7}%
   \end{figure*}
%   
%%%%% FIGURE 8 - Cumulatives2 %%%%% A figure as large as the width of the column %%%
   \begin{figure*}
   \centering
   \includegraphics[width=6cm,clip]{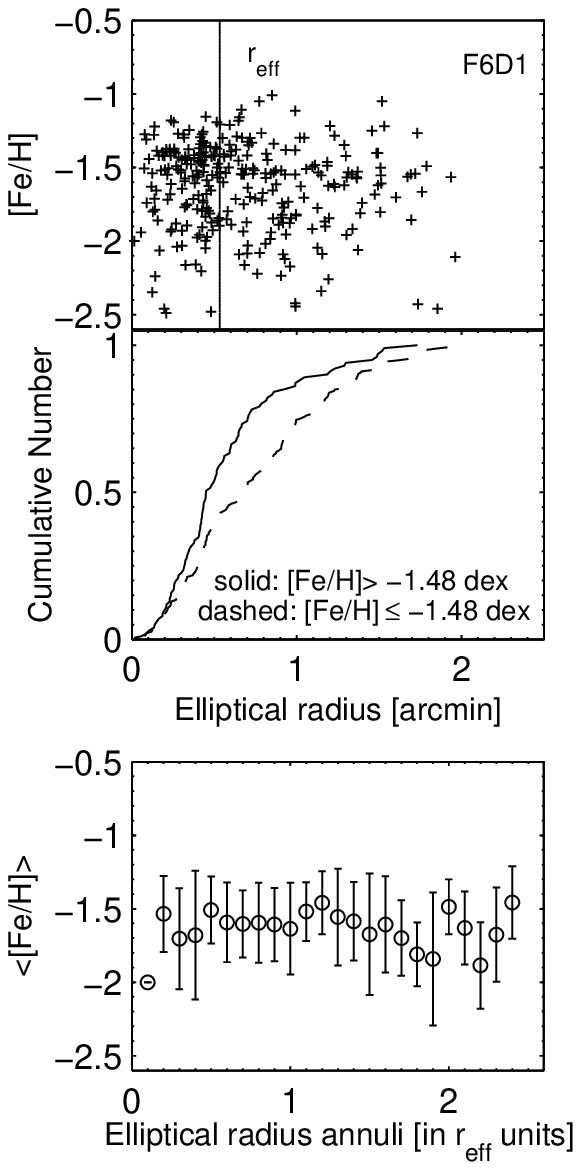}
   \includegraphics[width=6cm,clip]{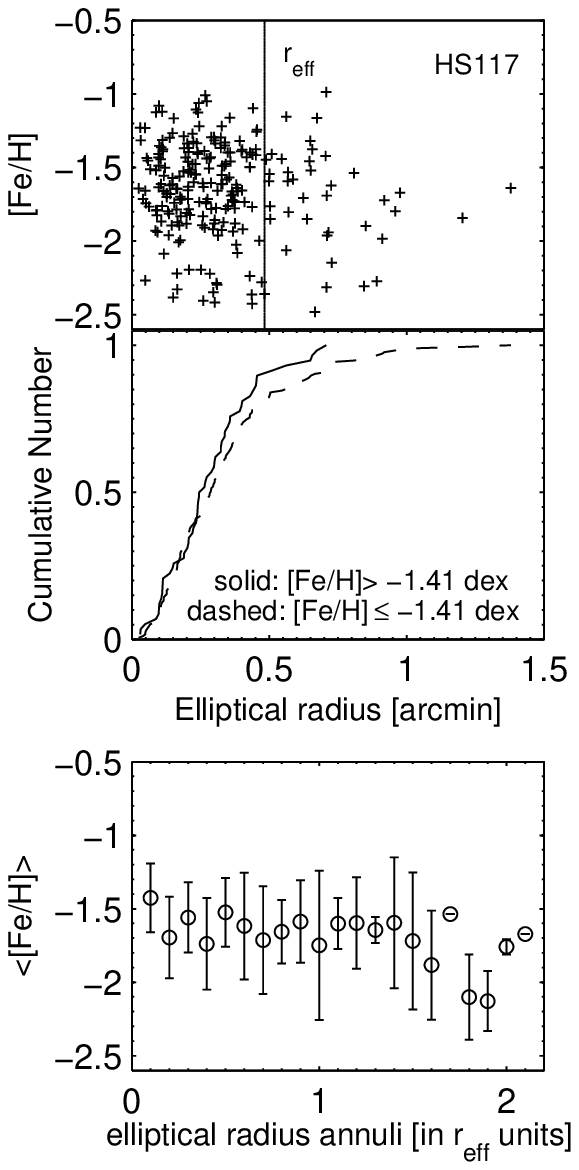}
      \caption{Same as in Fig.~\ref{sl_figure7} for the remaining two
               dSphs. Note that IKN is excluded from this analysis.}
   \label{sl_figure8}%
   \end{figure*}
%%%%%%%%%%%%%%%%%%%%%%%%%%%%%%%%%%%%%%%%%%%%%%%%%%%%%%%%%%%%%%%%%%%%%%%%%%%%%%%%%%%%
%
    We show the cumulative metallicity distributions in
    Fig.~\ref{sl_figure7} and Fig.~\ref{sl_figure8} (middle
    panels). We show in the same figures the radial metallicity
    distributions (upper panels) and the mean radial metallicity
    profiles (lower panels). Each radial metallicity profile shows the
    mean values of metallicity within an elliptical annulus versus the
    elliptical annulus in units of the effective radius $r_{eff}$. The
    values for $r_{eff}$ are listed in the column (10) of
    Table~\ref{table2}. The error bars in the metallicity profile
    correspond to the standard deviation of the mean metallicity in
    each elliptical radius annulus.

    \subsection{Density Maps}
    \label{sec:maps}

%%%%% FIGURE 9 - MAPS1 %%%%%%%%%%%% A figure as large as the width of the column %%%
   \begin{figure*}
   \centering
   \includegraphics[width=6cm,clip]{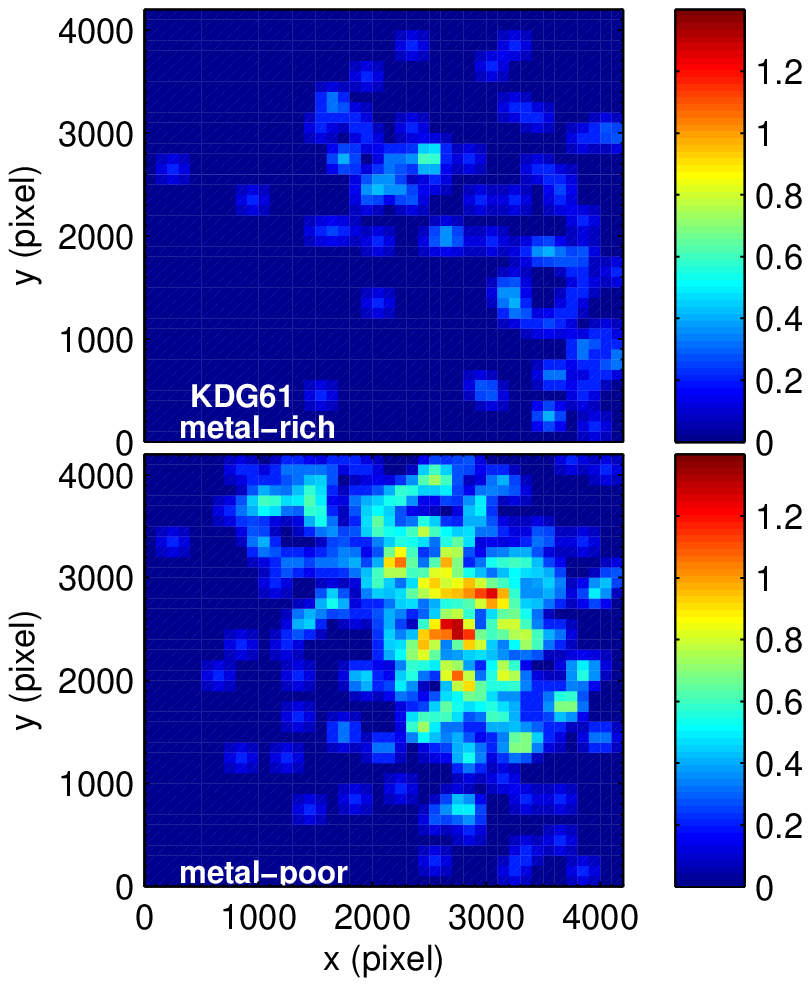}
   \includegraphics[width=6cm,clip]{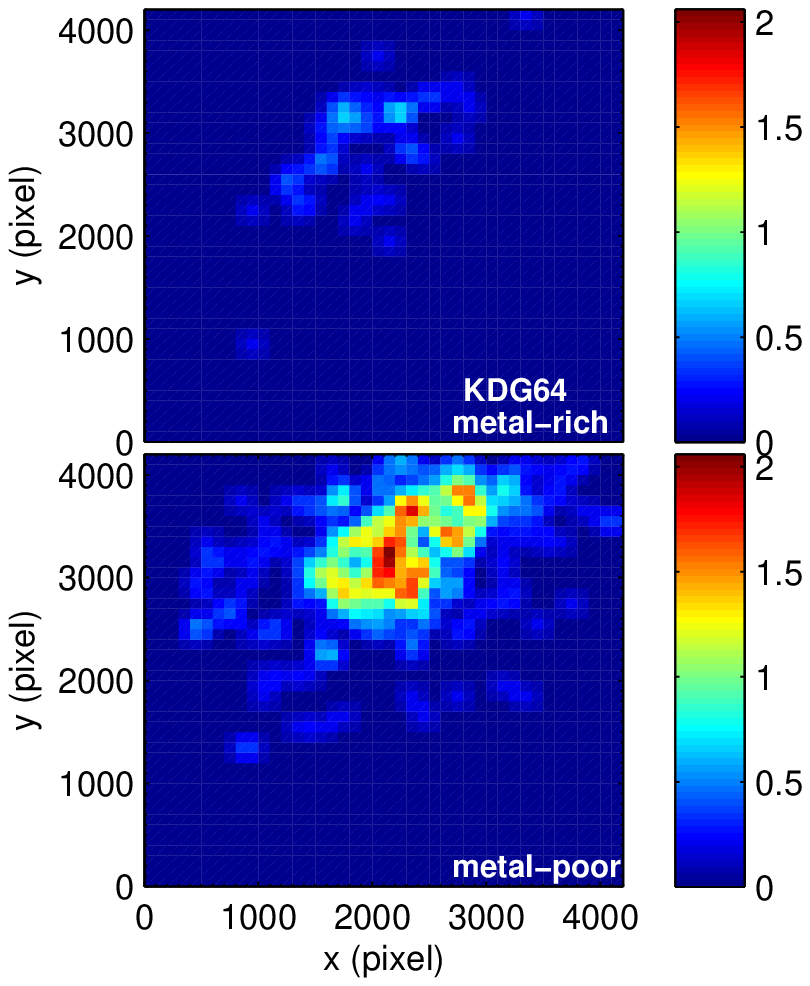}
   \includegraphics[width=6cm,clip]{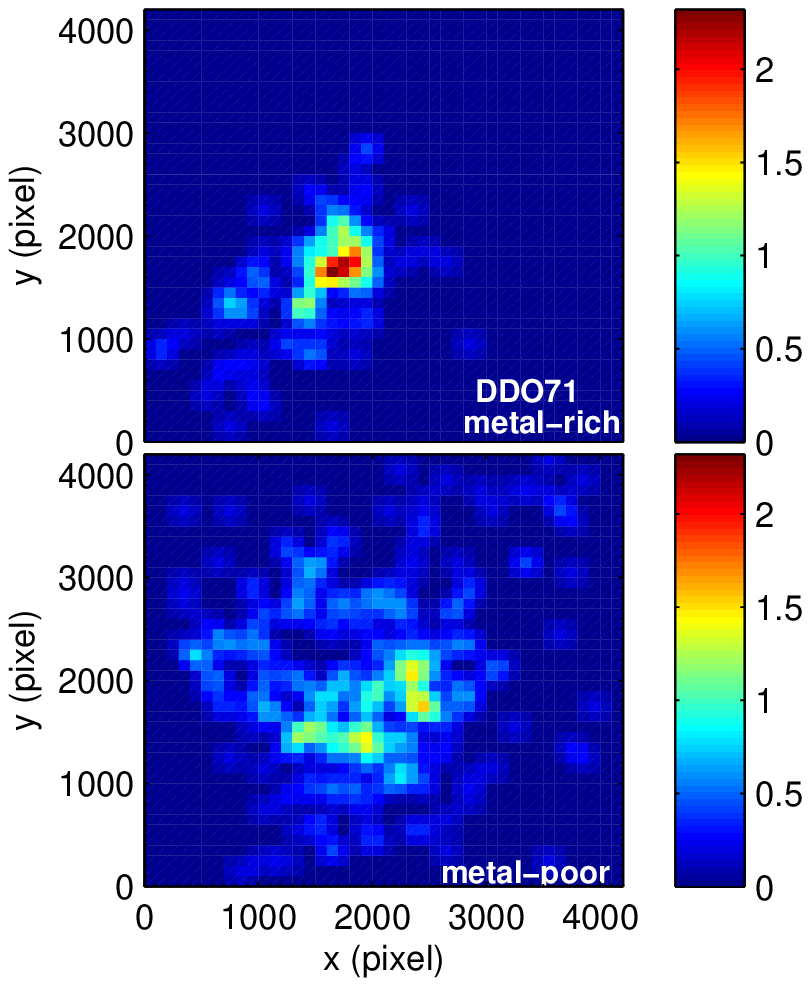}
   \includegraphics[width=6cm,clip]{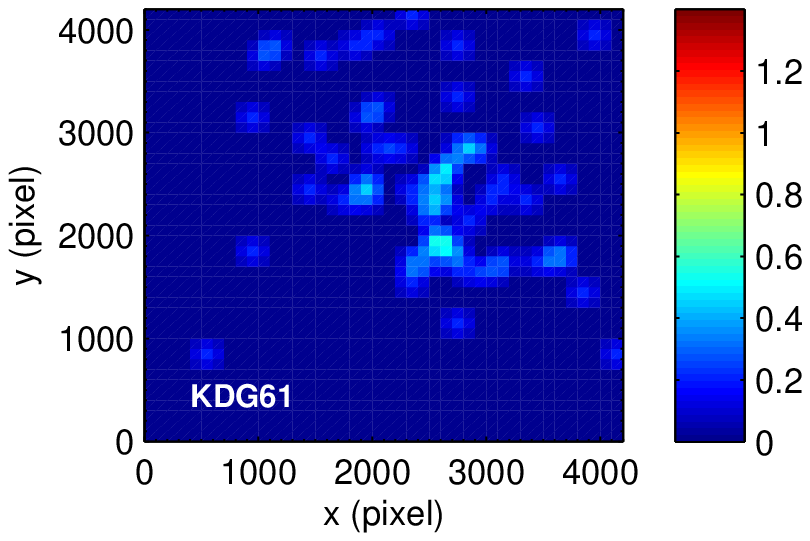}
   \includegraphics[width=6cm,clip]{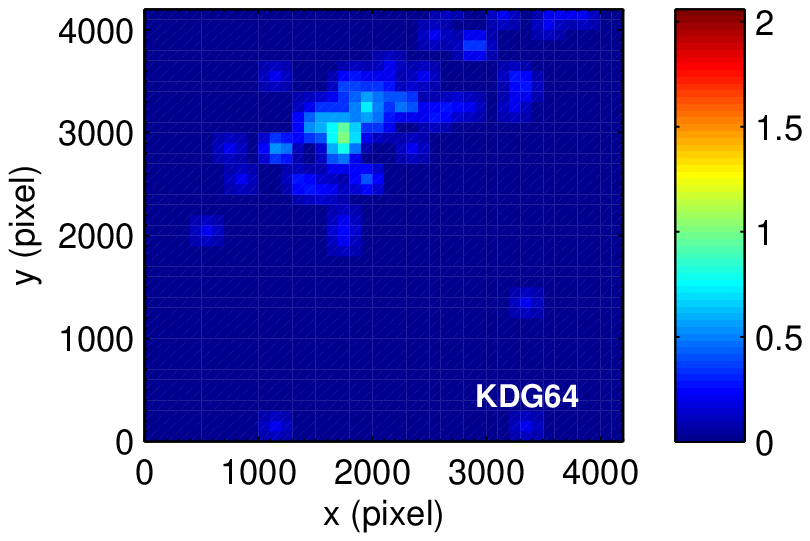}
   \includegraphics[width=6cm,clip]{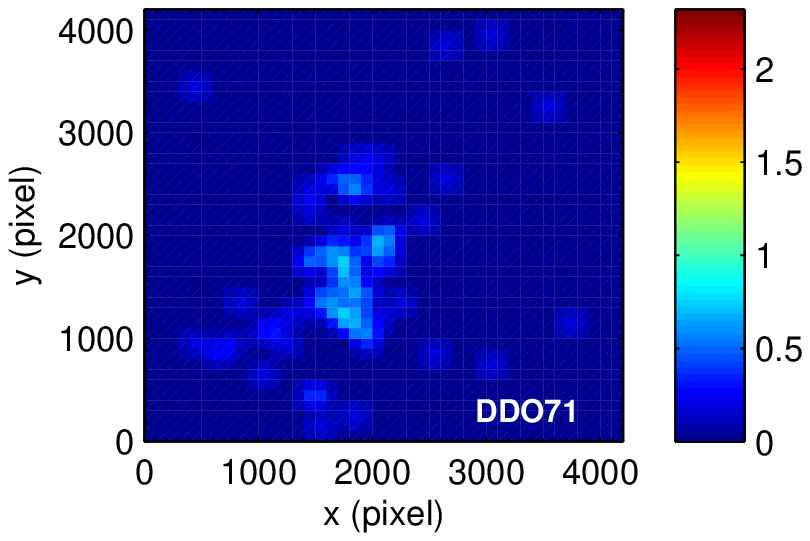}
   \includegraphics[width=6cm,clip]{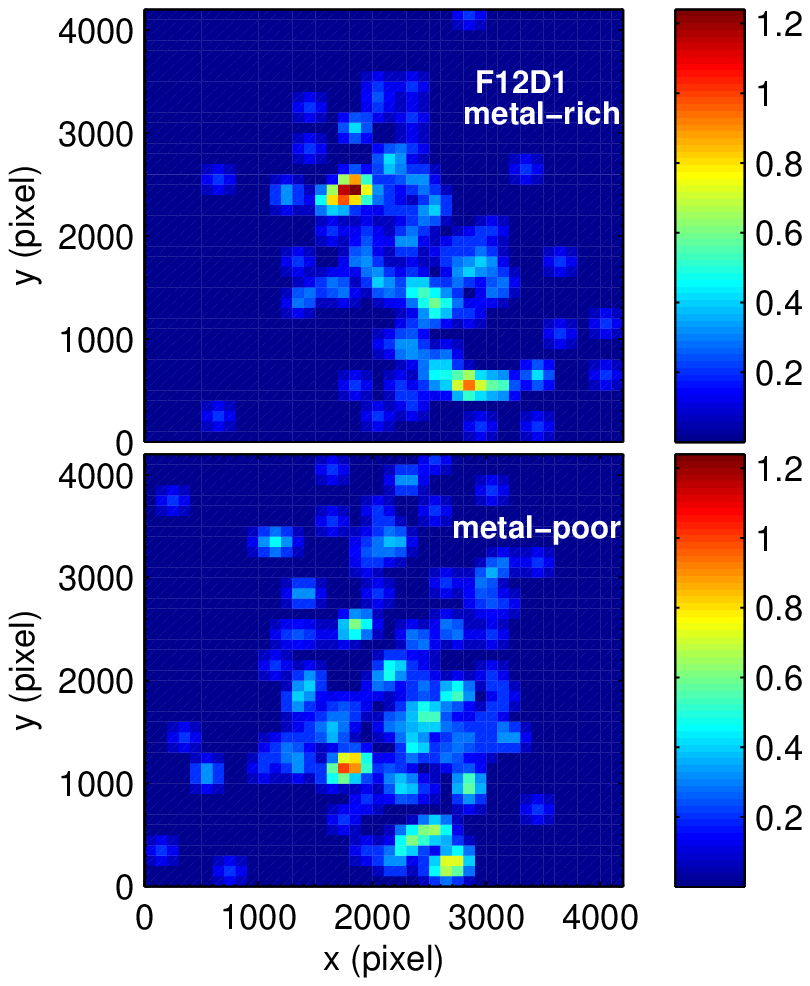}
   \includegraphics[width=6cm,clip]{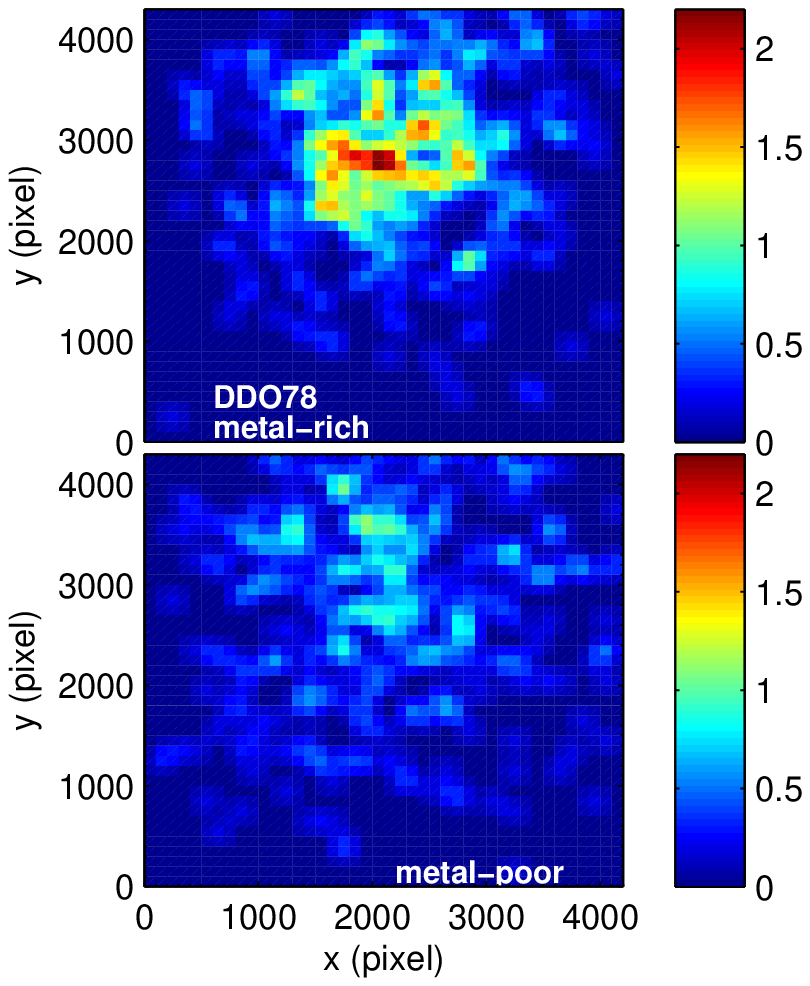}
   \includegraphics[width=6cm,clip]{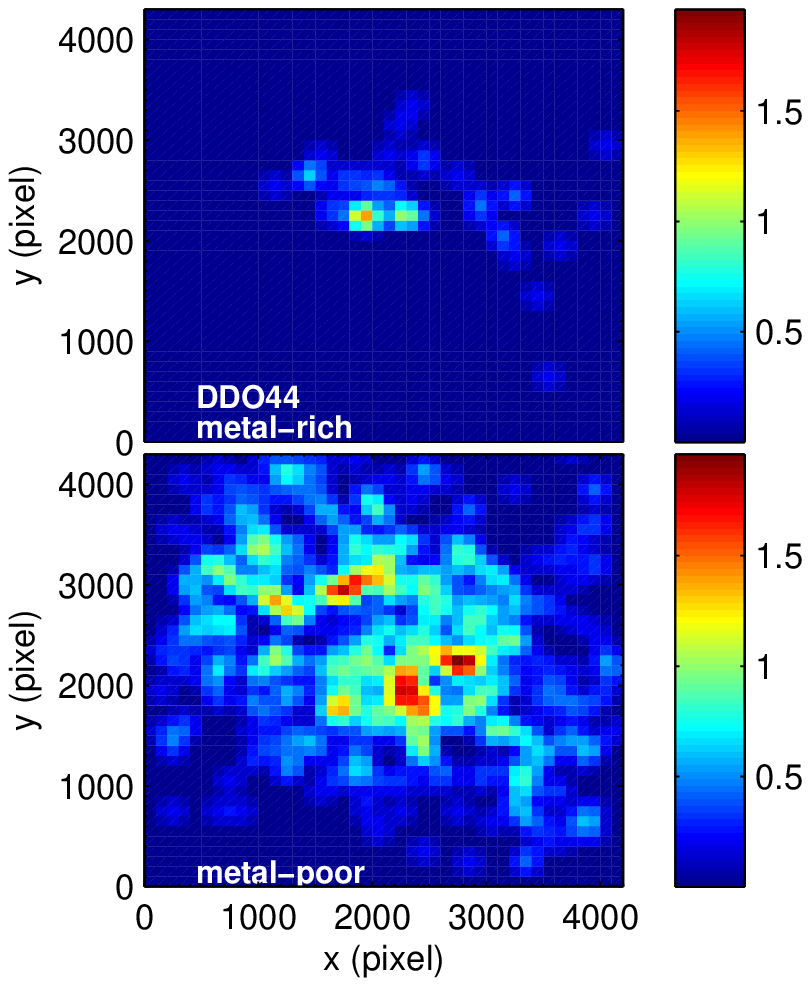}
   \includegraphics[width=6cm,clip]{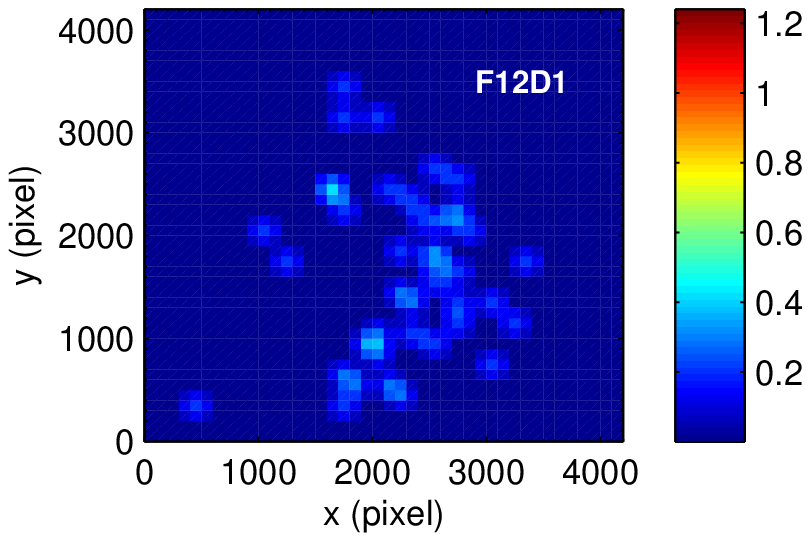}
   \includegraphics[width=6cm,clip]{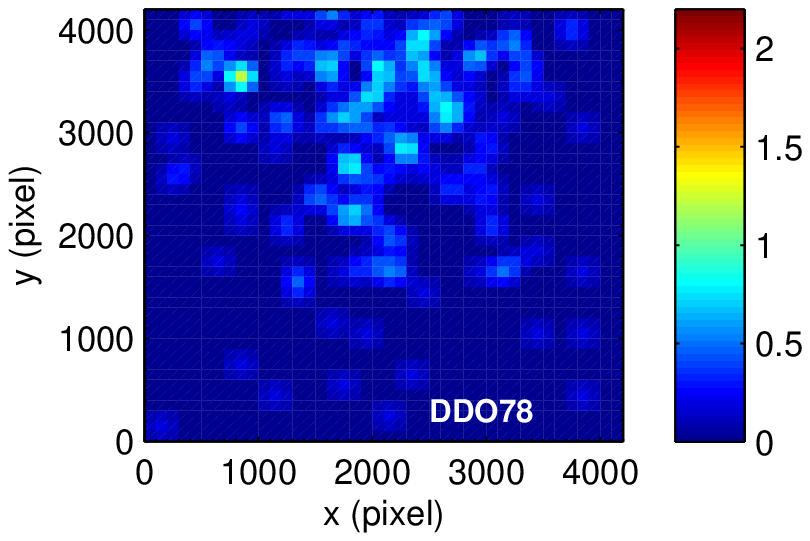}
   \includegraphics[width=6cm,clip]{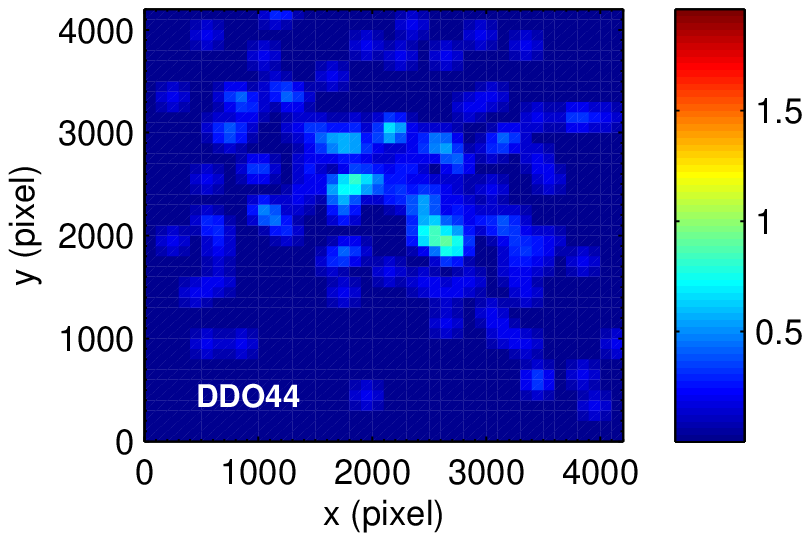}
     \caption{From left to right and from top to bottom we show the
              density maps for each dSph, after smoothing with a
              Gaussian kernel. In each upper and middle panel, the
              ``metal-rich'' population corresponds to stars having
              [Fe/H]$\ge-$1.30~dex while ``metal-poor'' refers to
              [Fe/H]$ \le-$1.80. The bottom panel corersponds to the
              density map of the luminous AGB stars, as defined in the
              text. The unit of the colorbars is number of stars per
              (100 pixels)$^2$.}
  \label{sl_figure9}%
  \end{figure*} 
%
%%%%% FIGURE 10 - MAPS2 %%%%%%%%%%%% A figure as large as the width of the column %%%
   \begin{figure*}
   \centering
   \includegraphics[width=6cm,clip]{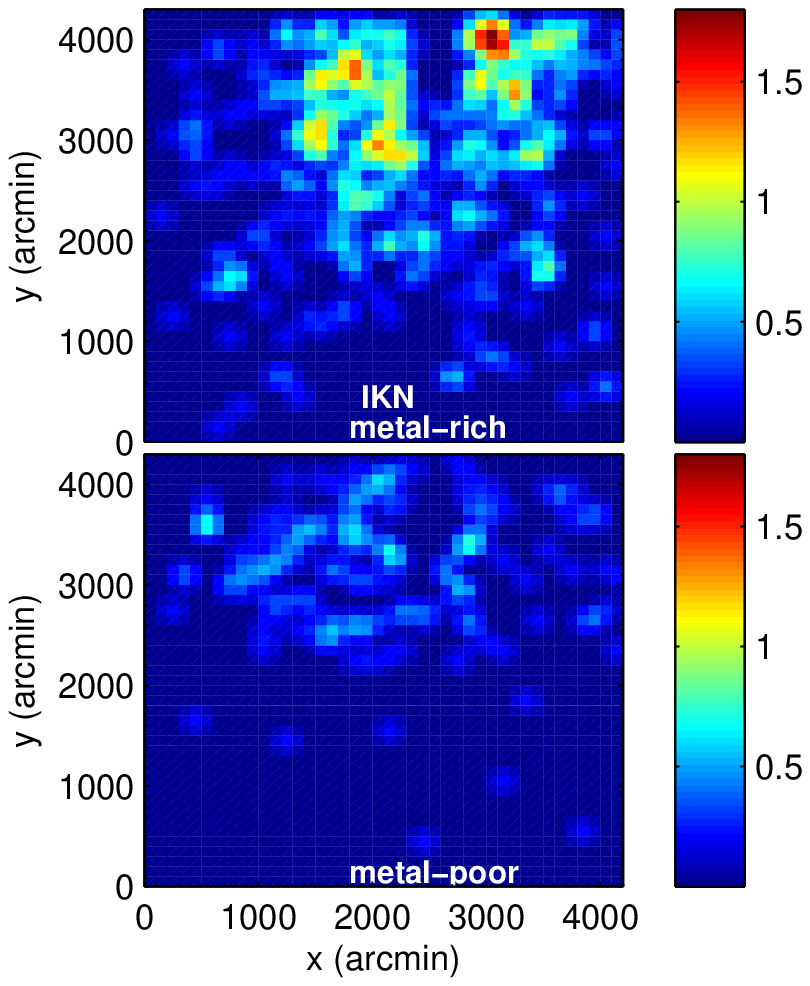}
   \includegraphics[width=6cm,clip]{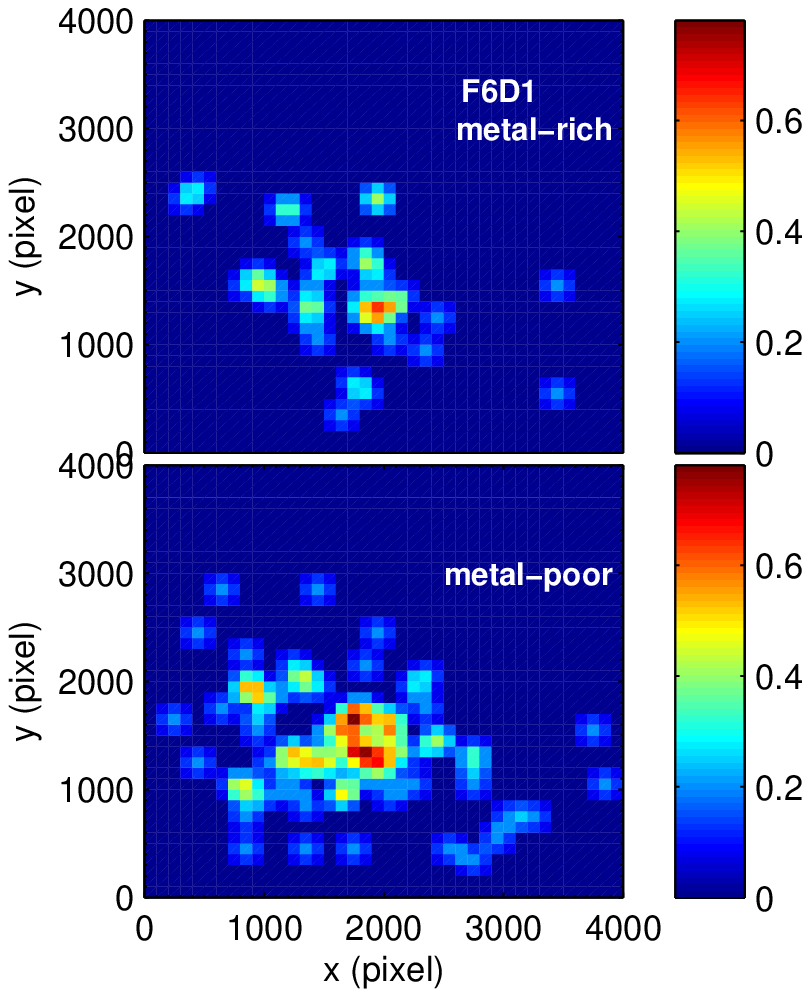}
   \includegraphics[width=6cm,clip]{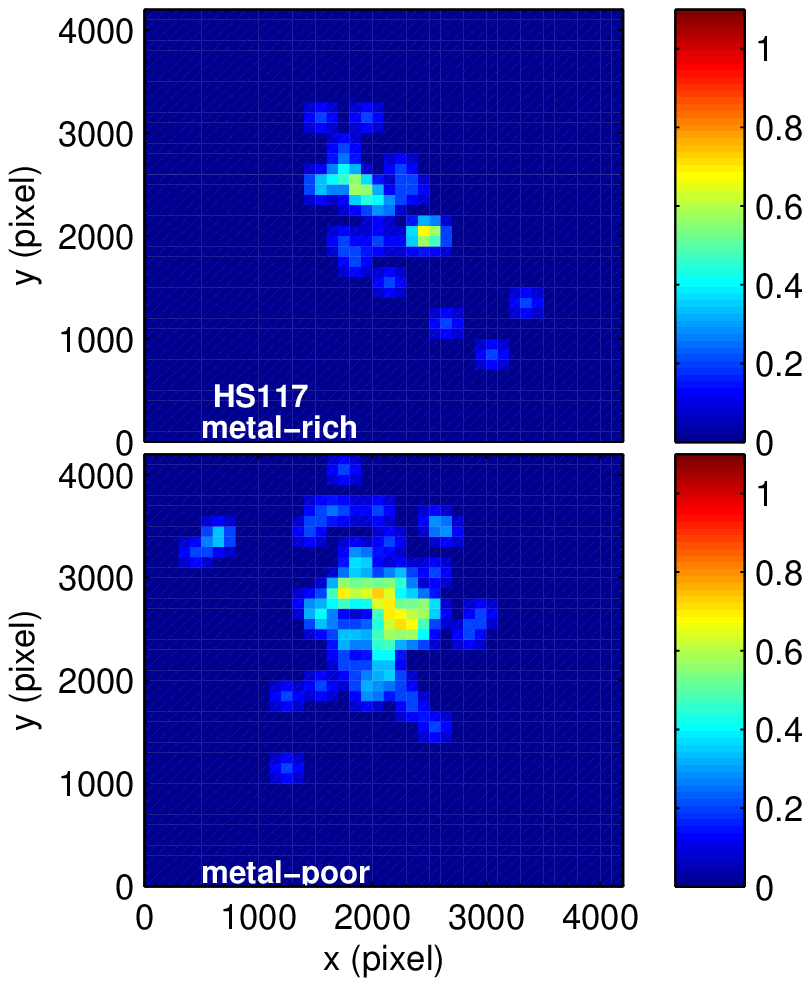}
   \includegraphics[width=6cm,clip]{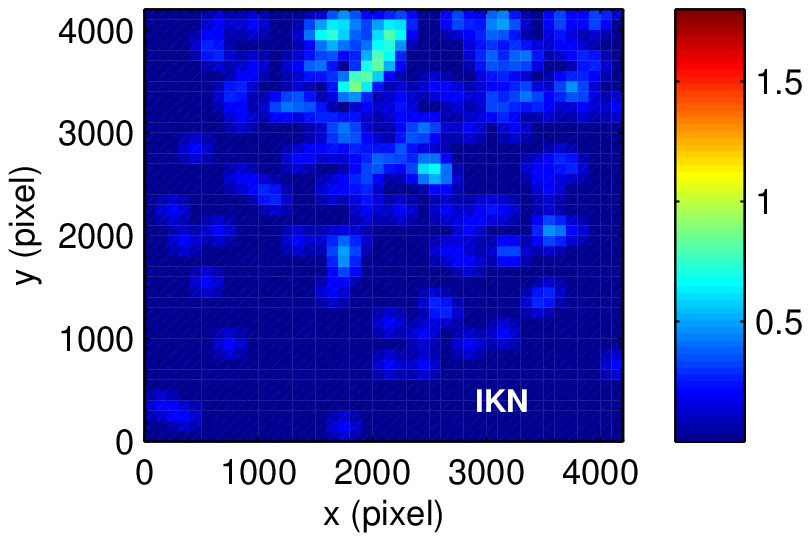}
   \includegraphics[width=6cm,clip]{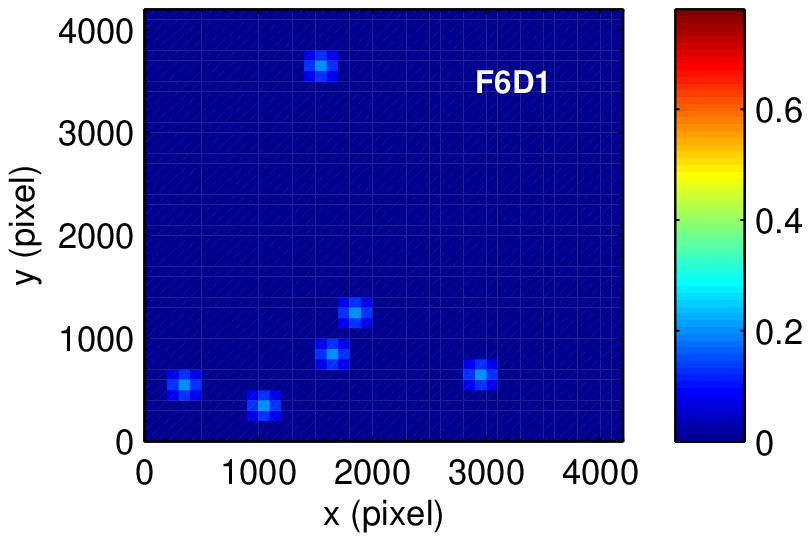}
   \includegraphics[width=6cm,clip]{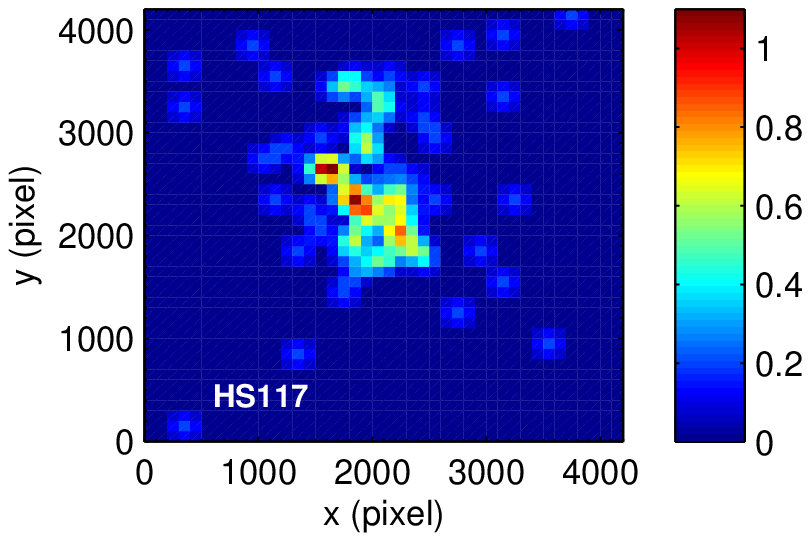}
      \caption{The same as in Fig.~\ref{sl_figure9} but for the
               remaining three dSphs. Note that the zero density
               region in the center of IKN is an artifact due to a
               bright star contaminating the field of view.}
   \label{sl_figure10}%
   \end{figure*}
%%%%%%%%%%%%%%%%%%%%%%%%%%%%%%%%%%%%%%%%%%%%%%%%%%%%%%%%%%%%%%%%%%%%%%%%%%%%%%%%%%%%
%
   We now examine the spatial distribution of the stellar populations,
   separated into a metal-poor and metal-rich component. For that
   purpose, we define two stellar populations, the first includes
   stars having a metallicity less than or equal to the value of
   $-$1.80~dex (''metal-poor''), while the second includes stars with
   a metallicity larger than or equal to $-$1.30~dex
   (''metal-rich''). All galaxies have peak values that lie well in
   between those cuts so that the metal-poor and the metal-rich tails
   are representatively sampled for all dSphs. For these two
   populations we construct the gaussian-smoothed density maps, shown
   in Fig.~\ref{sl_figure9} and  Fig.~\ref{sl_figure10}, upper and
   middle panels for each dSph.

\subsection{Luminous AGB stars}
\label{sec:lumagb}

    As already noted before, all the dSphs in our sample contain a
    significant number of luminous AGB star candidates. These stars
    are located above the TRGB and have ages ranging from 1~Gyr up to
    less than 10~Gyr. We broadly refer to stellar populations in this
    age range as ``intermediate-age'' populations. Assuming that the
    metallicities of dwarf galaxies increase with time as star
    formation continues, we may also assume that these
    intermediate-age populations are more metal-rich than the old
    populations.  We note, however, that dwarf galaxies do not
    necessarily experience smooth metal enrichment as a function of
    time (see, e.g., Koch et al.~\cite{sl_koch07a};
    \cite{sl_koch07b}).

    In Fig.~\ref{sl_figure9} and Fig.~\ref{sl_figure10}, the density
    maps in the lower panels show the spatial distribution of these
    intermediate-age stars for each dSph. We consider as luminous AGB
    stars the stars that are brighter by 0.15~mag than the TRGB
    (Armandroff et al.~\cite{sl_armandroff}) and that lie within 1~mag
    above ($I_{TRGB}-0.15$)~mag. In addition, we consider stars within
    the color range of $a\,<\,(V-I)_0\,<\,a+2.50$~(mag), where the
    left-hand limit $a$ is equal to the color of the TRGB of the most
    metal-poor isochrone we use, dereddened for each dSph using the
    extinction values listed in columns (6) and (7) of
    Table~\ref{table2}. Then, the right-hand limit is the left-hand
    limit plus 2.50~mag. This selection criterion was motivated by the
    work of Brewer, Richer \& Crabtree (\cite{sl_brewer95}) and Reid
    \& Mould (\cite{sl_reid84}).

\section{Discussion}
\label{sec:discussion}

    \subsection{Photometric Metallicity Distribution Functions}
    \label{sec:dmdfs}

    The photometric MDFs in Fig.~\ref{sl_figure2} indicate that these
    dSphs cover a wide range in metallicity. All of them seem to have
    a steeper cut-off in their metal-rich end. This can be easily seen
    if we compare the MDFs to the fitted gaussian distributions,
    indicated by the dashed lines in Fig.~\ref{sl_figure2}. We do not
    expect the MDFs to follow a gaussian distribution since they are
    shaped by the star formation histories of each dSph. For instance,
    a steep cut-off of the MDF toward the metal-rich tail could be
    indicative of the occurrence of strong and continuous galactic
    winds (Lanfranchi \& Matteucci \cite{sl_lanfranchi},
    \cite{sl_lanfranchi07}; Koch et al.~\cite{sl_koch06}) or of the
    effects of extended star formation and SNe Ia ``pockets'' of
    localized, inhomogeneous enrichment (Marcolini et al.~\cite{sl_marcolini08}).

    The low mean metallicities, $\langle$[Fe/H]$\rangle$, that are
    derived from the distribution functions and are shown in
    Table~\ref{table3}, columns (2) and (3), indicate that the M\,81
    dSphs are metal-poor systems, which points to a low star formation
    efficiency in analogy with the LG dSphs (e.g., Lanfranchi \&
    Matteucci \cite{sl_lanfranchi}; Grebel, Gallagher \& Harbeck
    \cite{sl_grebel}; Koch \cite{sl_koch09}).

    One exception is the dSph IKN, which shows a high mean metallicity
    for its luminosity. Objects that have high metallicity for their
    luminosity and that, most importantly, are dark matter free are
    promising candidates for tidal dwarf galaxies. Their properties
    are set by their formation mechanism. They are believed to form
    out of the dark matter free material that was expelled during the
    tidal interaction of the parent galaxies (Bournaud et
    al.~\cite{sl_bournaud}; and references therein). One possible
    tidal dwarf galaxy candidate has been identified in the M\,81
    group, namely Holmberg\,IX (Makarova et al.~\cite{sl_makarova};
    Sabbi et al.~\cite{sl_sabbi}) and these are favoured to be
    detected in such recently interacting groups. These systems
    contain a young stellar component, while their older stellar
    populations are believed to consist of stars from their parent
    galaxies, which is M\,81 in the case of Holmberg\,IX. There is no
    information available about the presence or absence of dark matter
    in this system.

    In the case of IKN, we should consider the fact that its stellar
    metallicity bears the imprint of the medium that formed these old
    stars, while young stars are not observed in this dwarf. That
    makes it distinct from young tidal dwarf candidates like
    Holmberg\,IX. A connection with the recent interactions of the
    M\,81 group is not obvious. IKN might be an old tidal dwarf galaxy
    if such systems exist. Alternatively, IKN may have undergone
    substantial mass loss in the past, leaving it as a low-luminosity
    but comparatively high-metallicity dSph. Without data on its
    detailed structure and kinematics, we cannot distinguish between
    these possibilities.

    The metallicity spreads of the studied dSphs are large, spanning
    1~$\sigma$ ranges from 0.27~dex to 0.37~dex, or intrinsic,
    error-weighted 1~$\sigma$ ranges from 0.16~dex to 0.43~dex. These
    abundance spreads are comparable to the ones observed in the LG
    dSphs (Grebel, Gallagher \& Harbeck \cite{sl_grebel}; Koch
    \cite{sl_koch09}) and may indicate the presence of multiple
    stellar populations and\,/\,or extended star formation
    histories. According to the models of Marcolini et
    al.~(\cite{sl_marcolini08}) and Ikuta \& Arimoto
    (\cite{sl_ikuta02}), the initial star formation in dSphs may have
    lasted as long as 3~Gyr or even longer, which can lead to a large
    dispersion in [Fe/H]. For ages older than 10~Gyr, the shape of the
    MDF does not depend strongly on age as described in
    Sec.~\ref{sec:mdfs} and shown in Fig.~\ref{sl_figure5}.

    \subsection{Luminosity-Metallicity Relation}
    \label{sec:dmetlum}

%%%%% FIGURE 11 - [Fe/H plots] %%%%% A figure as large as the width of the column %%%
   \begin{figure}
   \centering
   \includegraphics[width=5.5cm,clip]{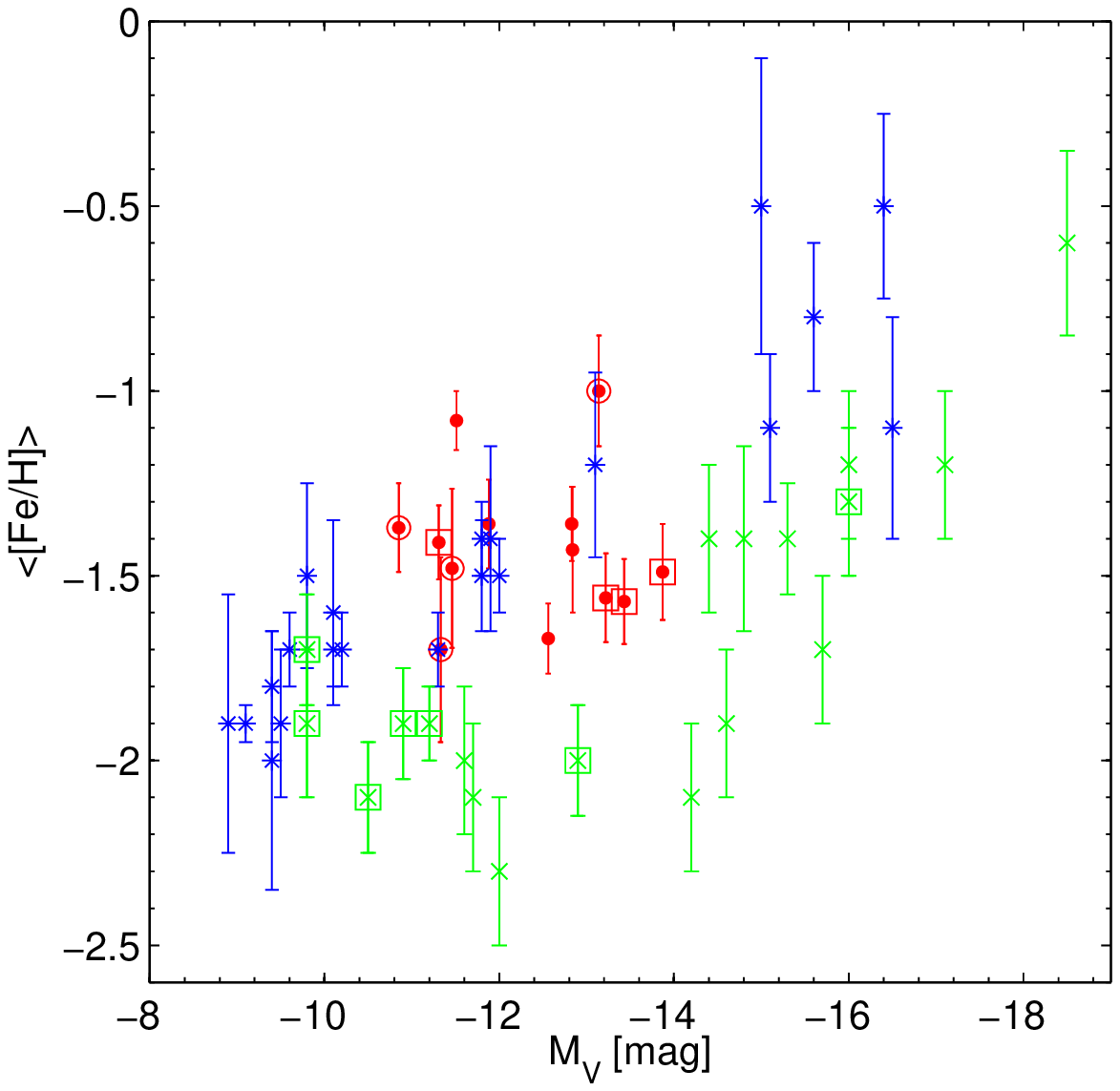}
   \includegraphics[width=5.5cm,clip]{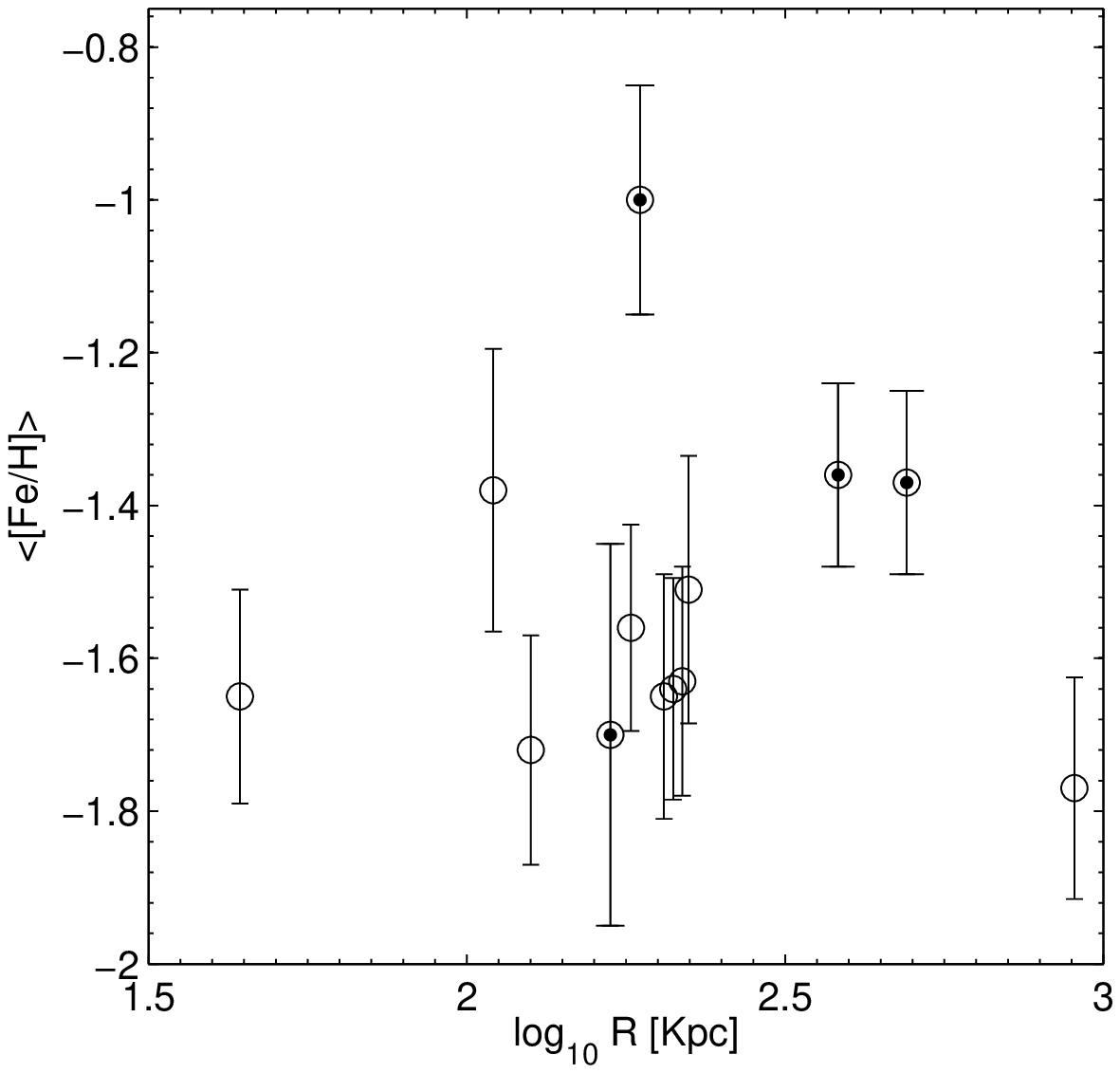}
   \includegraphics[width=5.5cm,clip]{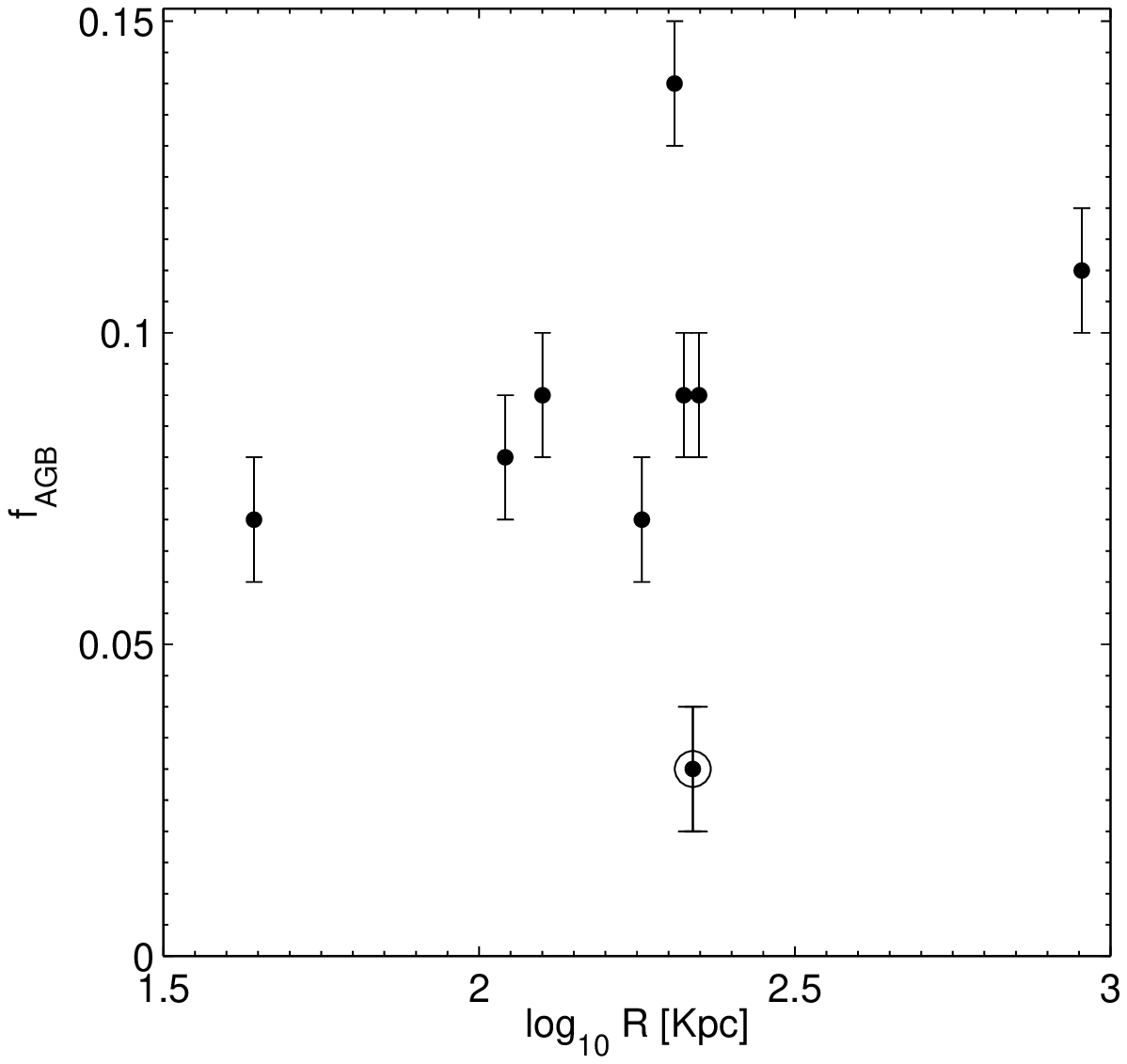}
      \caption{\textit{Upper}: Luminosity-metallicity relation for LG
               dwarf galaxies, after Grebel, Gallagher \& Harbeck\
               (\cite{sl_grebel}), together with the 13 dSphs of the M\,81
               group. LG dSphs are plotted with blue asterisks, LG dIrrs
               are shown as green crosses, and red dots indicate the
               available M\,81 data. Nine out of the thirteen M\,81 group
               dSphs have been studied here, while the remaining four,
               marked with a red circled dot, have been adopted from
               the literature, as discussed in the text. The red
               squared dots and green squared crosses indicate the
               transition-types of the M\,81 group and LG, respectively.
               \textit{Middle}: Mean metallicities versus the
               deprojected distance from M\,81, R, for the 13 M\,81 group
               dSphs. The circled dots correspond to the four dSphs
               for which the metallicities were adopted from the
               literature, as discussed in the text. 
               \textit{Lower}: Fraction of AGB stars versus the RGB
               stars within 1~mag below the TRGB, $f_{AGB}$, versus
               the deprojected distance from M\,81, R, for the nine M\,81 group
               dSphs studied here. With the circled dot we show the
               corresponding fraction $f_{AGB}$ in the case of F6D1.}
   \label{sl_figure11}%
   \end{figure}
%%%%%%%%%%%%%%%%%%%%%%%%%%%%%%%%%%%%%%%%%%%%%%%%%%%%%%%%%%%%%%%%%%%%%%%%%%%%%%%%%%%%
%
    In Fig.~\ref{sl_figure11}, upper panel, we show the
    luminosity-metallicity relation compiled for the dwarf galaxies in
    our LG as studied by Grebel, Gallagher \& Harbeck\
    (\cite{sl_grebel}), with the addition of thirteen dSphs in the
    M\,81 group. This compilation includes LG objects with mean
    metallicities derived from either spectroscopic or photometric
    studies. The mean metallicities for the nine M\,81 group dSphs
    listed in Table~\ref{table3}, column (3), are from this work,
    while the metallicities for the remaining four dSphs are adopted
    from the literature and are computed using the mean $(V-I)_{0}$
    color of the RGB stars at the luminosity of $M_{I}=-$3.5~mag (from
    Caldwell et al.~(\cite{sl_caldwell}) for BK5N and F8D1 and from
    Sharina et al.~(\cite{sl_sharina}) for KKH\,57 and BK6N).

    Overall, the M\,81 group dwarfs follow the luminosity-metallicity
    relation quite well, albeit some of them exhibit a tendency of
    being slightly more metal-poor than LG dSphs of comparable
    luminosity. Therefore, they mainly populate the region defined by
    the LG dSphs while a few are located in the border region between
    the dSph and dIrr locus defined by the LG dwarfs. The M\,81 group
    dSphs that seem to lie in this apparent transition region are
    KDG\,61, KDG\,64, DDO\,44 and DDO\,71. Out of these four objects,
    three are classified as transition-types, namely, KDG\,61, KDG\,64
    and DDO\,71 (Karachentsev \& Kaisin \cite{sl_m81halpha}; Boyce et
    al.~\cite{sl_hiblind}) based on HI detections and $H\alpha$
    emission. Also among the dwarfs that coincide with the LG dSph
    locus, one dwarf is classified as transition-type, namely HS\,117,
    with HI associated with it (Huchtmeier \& Skillman \cite{sl_huch};
    Karachentsev et al.~\cite{sl_acsdata}).

    Transition-type dwarfs are galaxies that share properties of both
    morphological types. Their stellar populations and star formation
    histories resemble those of dSphs and their gas content and
    present-day star formation activity is akin to low-mass dIrrs. It
    has been suggested that transition-type dwarfs are indeed evolving
    from dIrrs to gas-deficient dSphs.

    The projected spatial distribution of the dSphs within the M\,81
    group is shown in Fig. 1 of Karachentsev et
    al.~(\cite{sl_m81distances}), while their three-dimensional view
    is shown in their Fig.~6. In Fig.~\ref{sl_figure11}, middle panel,
    we plot the deprojected distances from the M\,81 galaxy, R, versus
    the mean metallicities for the 13 M\,81 group dSphs. The
    deprojected distances from the M\,81 galaxy, R, are adopted from
    Karachentsev et al.~(\cite{sl_m81distances}) and are listed in
    Table~\ref{table2}, column (9). The most distant dSph is DDO\,44
    which belongs to the NGC\,2403 subgroup, while KDG\,61 is the one
    closest to the M\,81 itself. Interestingely, the dwarf KDG\,61
    which is classified as a transition-type based on HI detections,
    is the closest to the M\,81 itself, with a deprojected distance of
    44~kpc (Karachentsev et al.~\cite{sl_m81distances}). The remaining
    three dwarfs classified as transition-types, namely KDG\,64,
    DDO\,71 and HS\,117, lie in a deprojected distance of more than
    100~kpc. As discussed already, the most metal-rich dSph studied by
    us is IKN, while according to the value that Caldwell et
    al.~(\cite{sl_caldwell}) provide for F8D1, this is the most
    metal-rich dSph in this group so far studied. We see no trend of
    the mean metallicity with the deprojected distance.

    \subsection{Population Gradients}
    \label{sec:dgradients}

    By examining the cumulative metallicity distributions in the
    middle panels in Fig.~\ref{sl_figure7} and Fig.~\ref{sl_figure8},
    we conclude that the metallicity gradients are present in the case
    of DDO\,71, DDO\,78, DDO\,44 and F6D1, while the metallicity
    gradients are less pronounced or not present in the remaining
    dSphs. We again separate the RGB stars in each dSph into two
    samples, where we choose to separate the distributions at the
    observed weighted mean metallicity. The probabilities from the
    two-sided Kolmogorov-Smirnov (K-S) test that the two components
    are drawn from the same parent distribution are listed in
    Table~\ref{table3}, column (4). The K-S results are consistent
    with showing spatially separated populations in the case of
    DDO\,71, DDO\,78, DDO\,44 and F6D1. In the case of the remaining
    dSphs, the gradients are less pronounced and the metal-rich and
    metal-poor populations have different distributions at the 84 --
    99.7~\% confidence level ($>$1.5~$\sigma$), except for HS\,117. In
    all cases, in which we observe such metallicity segregation, the
    sense is that more metal rich stars are more centrally
    concentrated, as also found in the majority of the LG dSphs
    (Harbeck et al.~\cite{sl_harbeck}; Tolstoy et
    al.~\cite{sl_tolstoy}; Ibata et al.~\cite{sl_ibata}).

    \subsection{Density Maps}
    \label{sec:dmaps}

    The density maps in Fig.~\ref{sl_figure9} and
    Fig.~\ref{sl_figure10} are useful to study the spatial
    distribution of the metal-rich (upper panels) and metal-poor
    (middle panels) population of each dSph. From these density maps
    we conclude that each dSph has a different stellar spatial
    distribution of their metal-rich and metal-poor stellar
    component. All of them show either a spatial variation of the
    centroids of the two stellar populations, as is the case of F12D1
    and KDG\,64, or that the metal-rich population is more centrally
    concentrated, as is the case of DDO\,71, DDO\,44 and F6D1. These
    findings agree well with the ones from the cumulative histograms,
    though we should keep in mind that the metal-poor and metal-rich
    stellar populations involved are differently selected. DDO\,78 and
    IKN are clearly fairly metal-rich while KDG\,61, KDG\,64 and
    DDO\,44 have prominent metal-poor populations.
 
    \subsection{Luminous AGB stars}
    \label{sec:dlumagb}

    Luminous AGB stars were also detected in two more dSphs in the M\,81
    group, namely BK5N and F8D1 (Caldwell et
    al.~\cite{sl_caldwell}). These luminous AGB stars may include
    carbon stars and may be long-period variables (LPVs) as have been
    found in other early-type dwarf galaxies (e.g., Menzies et
    al.~\cite{sl_menzies02}; Rejkuba et al.~\cite{sl_rejkuba06};
    Whitelock et al.~\cite{sl_whitelock09}), but we can not establish
    this for certain based on our current data. We compute the
    fraction of the luminous AGB stars, $f_{AGB}$, defined as the
    number of the luminous AGB stars, $N_{AGB}$, counted within the
    magnitude bin considered in Sec.~\ref{sec:lumagb}, over the number
    of the RGB stars within one magnitude below the TRGB,
    $N_{RGB}$. In order to estimate the $N_{RGB}$, we take into
    account that approximately 22~\% of the stars we count within one
    magnitude below the TRGB are old AGB stars (Durrell, Harris \&
    Pritchet \cite{sl_durrell01}). Thus, the $N_{RGB}$ is equal to
    78~\% of the stars we count within one mag below the TRGB. The
    fractions $f_{AGB}$ we derive in this way are listed in
    Table~\ref{table3}, column (5).

  \subsubsection{Blends, blue straggler progeny and foreground contamination}

    We now discuss the possibility that these luminous AGB stars may
    actually be (1) blends of bright RGB stars (Renzini
    \cite{sl_renzini98}), (2) blue straggler progeny (Guarnieri,
    Renzini \& Ortolani \cite{sl_guarnieri97}; and references
    therein), or (3) due to the foreground contamination.   

    In order to evaluate case (1), we use our artificial star
    experiments in order to quantify the number of the blends of
    bright RGB stars, $N_{blends}$, that may contribute to the
    detected number of the observed luminous AGB stars. We only
    consider here the case of a blend of two equal-magnitude
    stars. The magnitude of the blended star is always $0.75$~mag
    brighter than the initial magnitude of the two superimposed
    stars.

    We want to determine the location in the CMD of all the RGB stars
    that can end up as blends within the location in the CMD of the
    luminous AGB stars, as defined in Sec.~\ref{sec:lumagb} and
    further called as luminous AGB box. For that purpose, if we assume
    that the stars within the luminous AGB box were all blends, then
    they would originate from stars that have magnitudes 0.75~mag
    fainter. Thus, we shift the luminous AGB box by  0.75~mag towards
    the fainter magnitudes and furthermore we only consider the stars
    with magnitudes fainter than the TRGB. This procedure defines the
    location of the RGB stars that can end up as blends within the
    luminous AGB box and we call this the ``RGB blends box''.

    From the stellar catalogue we use as an input for the artificial
    star experiments, we select the stars that have such input
    magnitudes to place them within the ``RGB blends box''. From these
    input stars, we consider as blends the ones that have output
    magnitudes that can place them above the TRGB. We normalize the
    number of these blends to the number of the total input stars that
    are located within the ``RGB blends box''. Finally, the number of
    the observed blends is proportional to the number of the observed
    RGB stars located within the same ``RGB blends box''. Thus, the
    $N_{blends}$ for all the dSphs computed this way is equal to
    5~blends, 11~blends, 12~blends and 2~blends in the case of
    DDO\,44, DDO\,78, IKN and HS\,117, respectively, while in all the
    other cases the  number of blends is less than 1. Thus, the
    fraction of the blends defined as the number of blends divided by
    the number of the RGB stars within 1\,mag below the TRGB, is less
    than 0.6~\% in all cases but for IKN which is equal to 0.9~\%.  

    In the case (2), Guarnieri, Renzini \& Ortolani
    (\cite{sl_guarnieri97}) point out that the number of blue
    straggler progeny is of the order of $\sim$2~\% of all stars that
    reach the luminous AGB phase.

    In the case (3), we estimate the foreground contamination using
    the TRILEGAL code (Vanhollebeke, Groenewegen \& Girardi
    \cite{sl_trilegal1}; Girardi et al.~\cite{sl_trilegal2}). We count
    the number of foreground stars that fall within the same location
    in the CMD as in the luminous AGB box and these are considered to
    be the number of expected foreground contamination. In all the
    cases, the number of foreground stars is 4, with the exception of
    DDO\,71 and DDO\,44 where the number of foreground stars is 3 and
    5, respectively. This translates to a fraction of foreground
    stars, defined as the number of foreground stars divided by the
    number of RGB stars within 1~mag below the TRGB, of less than
    0.7~\%, with the exception of F6D1 where this fraction is equal to
    2~\%.

    We do not consider the case of old AGB LPVs, whose large amplitude
    variations may place them above the TRGB (Caldwell et
    al.~\cite{sl_caldwell}), as an additional source of contamination
    in the luminous AGB box, since the studied dSphs have mean
    metallicities of less than $-$1~dex. Such old AGB LPVs, with ages
    greater than or equal to 10~Gyr, have been observed above the TRGB
    in metal-rich ([Fe/H]\,$> -$1~dex) globular clusters (e.g.,
    Guarnieri, Renzini \& Ortolani \cite{sl_guarnieri97}; and
    references therein).

    We can now add all the contributions estimated in the above three
    cases, for each dSph. We call the sum of these three contributions
    the number of total contaminants, $N_{cont,tot}$, and compute
    their fraction $f_{cont,tot}\,=\,N_{cont,tot}\,/\,N_{RGB}$. The
    fraction of the total contaminants is less than 1~\% in all cases
    apart from IKN, HS\,117 and F6D1, where the fraction of the total
    contaminants is approximately 1.1~\%, 1.3~\% and 2.1~\%,
    respectively. We note that in all cases, there is a significant
    fraction of luminous AGB stars that can not be accounted for by
    considering the contribution of blends, binaries and foreground
    contamination. The dSph F6D1 is an exception to that, where the
    $f_{cont,tot}$ is $\sim$\,2~\%, as compared to the $f_{AGB}$ which
    is $\sim$\,3~\%. We note though that in the case of F6D1 the
    number of stars counted in the luminous AGB box is equal to 6
    stars. Thus, we conclude that the luminous AGB stars are a genuine
    population for the majority of the dSphs studied here.

  \subsubsection{Luminous AGB density maps and fractions}

    From the density maps of the luminous AGB stars shown in
    Fig.~\ref{sl_figure9} and  Fig.~\ref{sl_figure10}, lower panels,
    we see that if we consider the peak densities or the bulk of the
    luminous AGB stars, then it seems that these are more confined to
    the central regions of the dwarfs, a behaviour similar to what is
    found for the Fornax dSph (Stetson, Hesser \& Smecker-Hane
    \cite{sl_stetson}). If we consider the overall distribution then
    we note that for most of the dSphs these stars are rather more
    widely distributed, following the distribution of the metal-poor
    stars, with the exception of KDG\,64 and DDO\,71 where their AGB
    stars' distributions coincide mostly with the metal-rich
    population, which in the case of DDO\,71 is centrally
    concentrated. We conclude that the intermediate-age stellar
    component is well-mixed with the old stellar component.

    This behaviour is similar to the AGB stars' spatial distribution
    of the LG dwarfs. Indeed, Battinelli \& Demers
    (\cite{sl_battinelli04}; and references therein) discuss that for
    the LG dwarfs, for which there are carbon star studies, their
    Carbon-rich stars are distributed such that they coincide with the
    spatial distribution of the old stellar component. An exception is
    the dE NGC\,185, where the AGB stars are concentrated more in the
    centre than the old stellar component (Battinelli \& Demers
    \cite{sl_battinelli04b}), a similar behaviour as observed in the
    two M\,81 group dSphs discussed above.

    We plot the $f_{AGB}$ versus the deprojected distance from M\,81,
    R, in the lower panel of Fig~\ref{sl_figure11}. The highest
    fraction of luminous AGB stars is observed in HS\,117 and the
    lowest one in KDG\,61 and F12D1. We do not see any trend of the
    $f_{AGB}$ with increasing deprojected distance from M\,81. If we
    compute the net fraction of the luminous AGB stars, by subtracting
    the fraction $f_{cont,tot}$, due to the contribution of blends,
    binaries and foreground contamination, from the fraction $f_{AGB}$
    listed in the column (5) of Table~\ref{table3}, none of the trends
    and conclusions change.

\section{Conclusions}
\label{sec:conclusions}

    We use the CMDs of nine dSphs in the M\,81 group to construct
    their photometric MDFs. These MDFs show populations covering a
    wide range in metallicity with low mean metallicities indicating
    that these are metal-poor systems. All MDFs show a steeper
    fall-off at their high-metallicity end than toward their
    low-metallicity end indicating that galactic winds may play a role
    in shaping their distribution.  

    We compute the mean metallicity, $\langle$[Fe/H]$\rangle$, and the
    mean metallicity weighted by the metallicity error,
    $\langle$[Fe/H]$\rangle_{w}$, along with their corresponding
    standard deviations. The most metal-rich dSph in our sample is IKN
    even though it is the least luminous galaxy in our sample. IKN's
    comparatively high metallicity may indicate that it is a tidal
    dwarf galaxy or that it suffered substantial mass loss in the
    past. We do not see any correlation between the
    $\langle$[Fe/H]$\rangle$ and the deprojected distance from the
    M\,81 galaxy, R.

    We use the mean metallicity weighted by the metallicity errors,
    $\langle$[Fe/H]$\rangle_{w}$, to select two stellar populations
    having metallicities above and below that value. For these two
    stellar populations we construct cumulative histograms, as a way
    to search for population gradients in metallicity. We find that
    some dSphs show strong metallicity gradients, while others do
    not. In dSphs with radial metallicity gradients the more
    metal-rich populations are more centrally concentrated.

    Furthermore, we study the spatial (i.e., two-dimensional)
    distribution of our defined metal-rich and metal-poor stellar
    populations. This refined look no longer assumes radial symmetry,
    and we now find that in some dwarfs the metal-rich population is
    more centrally concentrated, while others show offsets in the
    centroid of the two populations. By examining the distribution of
    the luminous AGB stars, we conclude that, for the majority of the
    dSphs, these stars have mostly extended distributions, indicating
    that they have been well-mixed with the metal-poor stellar
    population. We do not find any correlation between the fraction of
    luminous AGB stars and the deprojected distance from the M\,81
    galaxy. While present-day distances may not be indicative of the
    dwarfs' position in the past and while their orbits are unknown,
    the apparent lack of a correlation between distance and
    evolutionary history may suggest that the evolution of the dwarfs
    was determined to a large extend by their internal properties and
    not so much by their environment. 

    Finally, there are some M\,81 dSphs that straddle the transition
    region between LG dSphs and dIrrs in the metallicity-luminosity
    relation. We may be observing low-luminosity transition-type
    dwarfs moving toward the dSph locus. Interestingly, these dwarfs
    are slightly more luminous than the bulk of the LG transition
    dwarfs. Perhaps some of the M\,81 dwarfs experienced gas stripping
    during the recent interactions between the dominant galaxies in
    the M\,81 group.

\begin{acknowledgements} The authors would like to thank an anonymous
                         referee for the thoughtful comments. We would
                         like to thank Rainer Spurzem and Thorsten
                         Lisker for useful discussions. SL and this
                         research were supported within the framework
                         of the Excellence Initiative by the German
                         Research Foundation (DFG) via the Heidelberg
                         Graduate School of Fundamental Physics
                         (HGSFP) (grant number GSC 129/1). SL would
                         like to acknowledge an EAS travel grant to
                         participate to the JENAM 2008 conference in
                         Vienna, where the preliminary results
                         presented here were shown. AK acknowledges
                         support by an STFC postdoctoral fellowship
                         and by the HGSFP of the University of
                         Heidelberg.

                         This research has made use of the NASA/IPAC
                         Extragalactic Database (NED) which is
                         operated by the Jet Propulsion Laboratory,
                         California Institute of Technology, under
                         contract with the National Aeronautics and
                         Space Administration. This research has made
                         use of NASA's Astrophysics Data System
                         Bibliographic Services. This research has
                         made use of SAOImage DS9, developed by
                         Smithsonian Astrophysical Observatory. This
                         research has made use of Aladin.

                         All of the data presented in this paper were
                         obtained from the Multimission Archive at the
                         Space Telescope Science Institute
                         (MAST). STScI is operated by the Association
                         of Universities for Research in Astronomy,
                         Inc., under NASA contract NAS5-26555. Support
                         for MAST for non-HST data is provided by the
                         NASA Office of Space Science via grant
                         NNX09AF08G and by other grants and
                         contracts.

\end{acknowledgements}

\end{document}